\documentclass[10pt,a4paper]{article}
\usepackage[margin=1in]{geometry}
\usepackage{placeins,float,amsmath,amssymb,booktabs,tabularx,array,enumitem,xcolor,hyperref,microtype,graphicx}
\usepackage{caption}
\hypersetup{hidelinks}

\newcolumntype{Y}{>{\raggedright\arraybackslash}X}
\title{SmartBloch-Agent: An Agentic Framework for Automated MRI Sequence Design and Optimization}
\author{
Jiakai Yang$^{1}$, Haoye Zheng$^{1}$, Yixiao Yue$^{1}$, Qinqin Yang$^{2}$,\\
Zhong Chen$^{1}$, Shuhui Cai$^{1,*}$, Congbo Cai$^{1,*}$\\
\small $^{1}$School of Electronic Science and Technology, Xiamen University, Xiamen 361005, China\\
\small $^{2}$Department of Radiological Sciences,\\
\small University of California Irvine, Irvine, CA 92697, United States of America\\
\small $^{*}$Corresponding authors: Shuhui Cai and Congbo Cai
}
\date{}
\begin{document}
\maketitle

\begin{abstract}
Magnetic resonance imaging (MRI) pulse sequences organize radiofrequency excitation, gradient encoding, and signal acquisition, determining magnetization evolution and image formation. Sequence research requires translating imaging objectives into executable protocols meeting physical and acquisition constraints, refined via physical simulation, signal reconstruction, and scientific evaluation. Sequence construction, physical verification, parameter search, and assessment involve many tools and manual decisions, limiting automation from objective to verifiable sequence. We develop SmartBloch-Agent, an integrated framework for automated MRI sequence design and optimization. Given objectives and scientific constraints in natural language, an AI agent interprets goals, plans tasks, orchestrates tools, and makes feedback-informed decisions, drawing on a construction module and optimizers for structural design and parameter search. Our GPU-accelerated SmartBloch engine supplies physical computation for both stages, so the framework configures, per sequence and objective, event validation, Full-Bloch magnetization evolution, signal acquisition, image reconstruction, and scientific evaluation, and closes the loop without external MRI simulators. Results return to the agent to guide sequence revision, parameter optimization, and candidate selection, and the design files carry simulation, reconstruction, and evaluation evidence, forming an output package for a real-scanner workflow. Experiments show the framework automatically designs and optimizes MRI sequences for varied objectives and constraints, refines candidates with physical feedback, and improves several parameter-estimation metrics and downstream quantitative reconstruction. SmartBloch-Agent unifies objectives, design, physical execution, and evaluation into a continuous workflow for end-to-end automation of MRI sequence research.
\end{abstract}

\section{Introduction}\label{sec:introduction}
Magnetic resonance imaging (MRI) uses radiofrequency (RF) excitation, gradient fields, and signal acquisition to measure tissue structure and physical properties noninvasively. Pulse sequences define the temporal organization and execution of these events. Sequence design directly affects image contrast, spatial resolution, acquisition efficiency, and quantitative information, making sequence research an important driver of advances in MRI. Bloch's 1946 description of nuclear magnetization dynamics provided a physical foundation for predicting magnetization evolution~\cite{bloch1946induction}. In 1973, Lauterbur demonstrated spatial magnetic resonance imaging with spatially varying magnetic fields~\cite{lauterbur1973imaging}. Controlling magnetization and acquiring its resulting signals subsequently became central problems in sequence research.

As MRI evolved from basic imaging toward rapid acquisition and complex encoding, sequence research expanded the range of possible measurements. Small-flip-angle rapid imaging methods developed in the 1980s enabled acquisitions with short repetition times~\cite{haase1986flash}. In the late 1990s, parallel imaging further improved acquisition efficiency by exploiting the spatial sensitivities of multiple receiver coils~\cite{pruessmann1999sense}. These advances arose not only from hardware improvements but also from coordinated developments in excitation, echo formation, spatial encoding, and reconstruction. Researchers increasingly had to manage complex event timing, hardware limits, magnetization responses, and image reconstruction. Sequence research thus evolved from constructing individual pulse protocols into an integrated activity involving physical modeling, computation, and system implementation.

This study addresses two related tasks in MRI sequence research: \emph{sequence design} and \emph{sequence optimization}. Sequence design establishes or modifies the organization of RF, gradient, timing, and readout events for a specified imaging objective and converts the resulting protocol into an executable sequence. Sequence optimization searches a defined space of parameters or waveforms under physical constraints to improve contrast, acquisition efficiency, signal characteristics, or quantitative performance. Design determines which measurement mechanisms and sequence structures can be explored, whereas optimization identifies suitable operating conditions. Although these tasks act on different variables, both require physical simulation, signal reconstruction, and scientific evaluation to assess candidate sequences.

MRI sequence research has traditionally been coordinated by domain experts. Researchers develop protocols from physical principles and imaging requirements, write sequence programs, check the validity of gradient and RF events, inspect signals through Bloch simulations or scanner experiments, and repeatedly revise their designs based on reconstructed images and quantitative measures. This process requires combined expertise in pulse sequences, magnetization dynamics, computational models, and experimental conditions. A change in sequence structure or parameters may alter event timing, physical response, and reconstruction performance simultaneously, requiring renewed verification. The translation of research goals into specifications, integration of simulation and optimization tools, interpretation of evaluation results, and choice of subsequent revisions often remain manual. As the number of variables and constraints increases, cross-stage coordination and repeated computation become substantial components of research cost.

Computational tools have gradually changed this process. Numerical MRI simulators and portable sequence representations make candidate protocols easier to execute and compare, enabling some physical checks before scanner experiments~\cite{stocker2010jemris,layton2017pulseq,castillopassi2023komamri}. Magnetic resonance fingerprinting (MRF), introduced in 2013, demonstrated the role of acquisition timing in encoding quantitative information~\cite{ma2013mrf}. Subsequent optimal experimental design based on the Cramér--Rao lower bound (CRLB) linked spin dynamics to parameter-estimation precision~\cite{zhao2019oedmrf}. In the 2020s, differentiable MRI simulators and automatic differentiation allowed sequence events and RF and gradient waveforms to be optimized for specified objectives~\cite{loktyushin2021mrzero,luo2021autodiff,endres2024pdg}. Derivative-free methods have also used measured scanner feedback to explore imaging contrast automatically~\cite{glang2022mrdoublezero}. These studies have shifted candidate generation and parts of evaluation from manual trial and error toward automated computation. A complete research task, however, still requires objective interpretation, methodological choices, integration of tools, and adaptation of the research strategy to experimental feedback.

These requirements create an opportunity for AI agents in MRI sequence research. Unlike algorithms devoted to solving a predefined optimization problem, an agent can coordinate multiple steps around a research objective, invoke sequence construction, physical simulation, numerical search, and evaluation tools, and decide what to do next based on obtained results. Tool use and feedback-driven decisions have been investigated systematically in approaches such as ReAct~\cite{yao2023react}, while Coscientist illustrated the potential of language models for scientific experiment planning and tool coordination~\cite{boiko2023coscientist}. In MRI, GPT4MR explored language-model assistance for sequence and reconstruction programming~\cite{zaiss2024gpt4mr}. The 2026 arXiv preprint Agent4MR further combined executable sequence code, physical verification, and simulation feedback for automated development and refinement~\cite{zaiss2026agent4mr}. Together, these developments show that agents can participate in program generation and iterative decisions in sequence research and motivate the integration of autonomous decisions with reliable physical computation, numerical optimization, and scientific evaluation.

The central challenge in fully automated sequence research is to establish a continuous process connecting research objectives, executable sequences, and scientific outcomes. Candidate generation is only one part of that process. Natural-language objectives must be converted into explicit design requirements, adjustable variables, and constraints. Candidate events must satisfy physical and timing requirements before expensive magnetization evolution and reconstruction calculations. Different research tasks require different performance measures, and multiobjective results need consistent evaluation criteria. An automated workflow must also use failure reports and quantitative feedback to determine when to revise structures, adjust parameters, continue computations, or select a final design. Flexible agent decisions therefore need to operate together with executable physical models and deterministic scientific evaluation throughout candidate generation, verification, and selection.

We therefore propose SmartBloch-Agent, a fully automated framework for MRI sequence design and optimization. Given research objectives and required constraints, an agent manages task interpretation, research planning, tool coordination, and feedback-informed decisions. Sequence construction and numerical optimization tools generate candidates, while Bloch simulation, reconstruction, and scientific evaluation guide successive iterations. Our contributions are threefold. \textbf{(1) Automated MRI sequence design} connects research requirements to executable sequence construction, physical verification, and feedback-based revision. \textbf{(2) Automated MRI sequence optimization} combines agent decisions with numerical search and scientific metrics for iterative improvement based on physical feedback. \textbf{(3) The SmartBloch simulation engine}, developed in-house, accelerates Bloch magnetization evolution using CUDA parallel computing and provides internal physical computation for both tasks. Together, these components form an automated workflow from research objectives to sequence candidates, physical evidence, and study outcomes, with execution, acquisition, reconstruction, and evaluation configured for the sequence type and research objective.

\section{Related Work}\label{sec:related}
\subsection{MRI Sequence Design}
Sequence development translates imaging objectives into RF, gradient, and readout events while respecting magnetization dynamics, hardware amplitude limits, and timing constraints. Shared sequence representations are an important means of reducing development effort. Pulseq stores events and timing in standalone files interpreted by the scanner, separating prototype development from vendor-specific programming environments~\cite{layton2017pulseq}. Pulseq-GPI provides a visual programming environment integrated with simulation and reconstruction~\cite{ravi2018pulseqgpi}, and PyPulseq brings sequence construction into Python workflows~\cite{ravi2019pypulseq}. These tools support development and exchange through common event representations, with automation mainly at the representation and execution interfaces.

Automated generation extends computation to sequence events themselves. MRzero constructs an end-to-end mapping through differentiable magnetization evolution and reconstruction. It adjusts RF events, gradient moments, and delays for target images or parameter maps and validates generated sequences in phantom and in vivo measurements~\cite{loktyushin2021mrzero}. Its task-loss-driven search extends beyond manual tuning of an existing protocol. A separate approach uses language models to generate programs. GPT4MR assists sequence and reconstruction programming through domain-specific prompting and interactive feedback~\cite{zaiss2024gpt4mr}. SeqGPT trains a generative model on pseudorandom sequences and reports the generation of simulatable sequence structures~\cite{hussain2025seqgpt}.

Programming tools provide event representations, objective-driven methods search events or parameters, and language-model generators produce executable programs. Agent workflows additionally coordinate verification and revisions. Our framework places structural design and parameter optimization within a common research process, using sequence-specific execution modules to turn candidates into physically evaluable sequences.

\subsection{MRI Sequence Optimization}
Sequence optimization expresses imaging requirements as computational objectives and searches for parameters subject to acquisition constraints. By encoding tissue properties with varying acquisition patterns, MRF~\cite{ma2013mrf} motivated a shift from signal amplitude toward the design of quantitative information. Comparisons of optimization algorithms examined the effects of search strategies on acquisition protocols~\cite{cohen2017algorithms}. CRLB-based optimal experimental design connects spin dynamics, noise assumptions, and parameter-estimation uncertainty~\cite{zhao2019oedmrf}. B-spline parameterization represents time-varying acquisition patterns with a small number of coefficients, reducing a high-dimensional design to a compact search~\cite{scopecrafts2022bsplinemrf}. Studies of encoding capacity and parameter-map error have examined noise propagation and aliasing, respectively, informing the selection of evaluation objectives~\cite{sommer2017encoding,kara2019maperror}.

Differentiable methods propagate gradients through physical models. MRzero jointly considers sequence events and reconstruction, allowing target-image error, scan time, and RF energy requirements to enter the same optimization process~\cite{loktyushin2021mrzero}. For finer waveform design, automatic differentiation supports joint optimization of RF and gradient waveforms for tailored three-dimensional excitation~\cite{luo2021autodiff}. AutoCEST links saturation modules, spin dynamics, and deep reconstruction networks in a differentiable system, jointly learning acquisition protocols and molecular parameter estimates, with validation in phantoms and mouse brains~\cite{perlman2022autocest}. These methods use task losses and model structure to guide the joint design of acquisition and reconstruction.

Physical feedback can also come directly from a scanner. MR-double-zero uses derivative-free CMA-ES search and feedback based on the regression error between measured images and a target to adjust preparation-pulse parameters automatically in chemical-exchange contrast experiments~\cite{glang2022mrdoublezero}. This demonstrates contrast discovery without an explicit target signal model. Model-based gradients, black-box numerical search, and measured feedback suit different research tasks. An agent can select among tools, interpret evaluations, and coordinate subsequent iterations. SmartBloch-Agent retains the complementary roles of numerical search and physical evaluation and coordinates candidate generation and multiobjective selection at the workflow level.

Optimization methods may modify acquisition parameters, RF and gradient waveforms, or reconstruction strategies. They provide complementary search capabilities, while task decomposition and the coordination of physical execution with multiobjective feedback remain workflow-level problems.

\subsection{MRI Physical Simulation and High-Performance Computing}
The Bloch equations describe magnetization evolution under applied fields and relaxation~\cite{bloch1946induction}. Applying them to sequence research requires modeling event timing, spatial encoding, tissue distributions, and received signals rather than relying solely on simplified signal expressions. JEMRIS supports three-dimensional simulations through modular sequences and parallel computation, including multichannel transmission and reception, off-resonance, nonlinear gradients, and time-varying parameters~\cite{stocker2010jemris}. Such event-level simulation links candidate events to reconstructed data while making computational cost dependent on tissue sampling and model complexity.

GPU platforms exploit parallel spin evolution to increase simulation scale, although their modeling priorities differ. MRiLab incorporates a generalized multipool exchange model and uses GPU computation for exchange- and magnetization-transfer-related simulations~\cite{liu2017mrilab}. BlochSolver executes experimentally compatible sequences, reproduces images through subvoxel sampling, and studies the effects of sampling density on artifacts and computational cost~\cite{kose2017blochsolver}. KomaMRI.jl provides CPU/GPU Bloch solvers, Pulseq-compatible input, and reconstruction interfaces in Julia~\cite{castillopassi2023komamri}. GPU-JEMRIS subsequently moved numerical integration onto GPUs and evaluated single- and double-precision performance over different spin populations~\cite{nurdinova2025gpujemris}.

Another approach to acceleration changes the representation of magnetization states. Phase distribution graphs (PDGs) describe dephasing states following the extended-phase-graph concept, incorporate spatial encoding and arbitrary timing into differentiable computation, and support gradient-based optimization of RF angles, phases, gradients, and timing~\cite{endres2024pdg}. This analytical state representation differs computationally from sampling large spin ensembles; the implementation assumes instantaneous, on-resonance RF pulses. Automated research repeatedly invokes physical evaluators, so it depends on both computational throughput and consistency between the model and study objectives. We integrate SmartBloch as an internal Bloch simulation engine for different candidate-generation pathways, with execution and evaluation configured to each sequence and objective.

\subsection{Scientific Agents and Automated MRI Research}
Scientific automation has expanded from repetitive procedures to experimental selection and learning from feedback. Self-driving laboratories integrate robotic synthesis, measurements, and Bayesian optimization to identify thin-film processing conditions automatically~\cite{macleod2020selfdriving}. Multiobjective extensions explore Pareto trade-offs between electrical conductivity and processing temperature~\cite{macleod2022pareto}. These studies show that closed-loop experimental automation predates language-model agents and fundamentally depends on connecting objectives, executable actions, and measured feedback.

Language models extend task planning across tools. ReAct updates plans by alternating reasoning, action, and environmental observations~\cite{yao2023react}. Coscientist integrates retrieval, code execution, and experimental equipment for planning and performing chemistry experiments and optimizing with existing data~\cite{boiko2023coscientist}. The formally published 2026 paper on The AI Scientist organizes idea generation, coding experiments, result analysis, and paper writing in machine-learning research~\cite{lu2026aiscientist}. These works offer evidence for general interaction mechanisms, coordination of physical experiments, and computational research workflows, respectively.

Agent4MR is the most directly related work in MRI. This preprint connects PyPulseq code generation and execution, timing and trajectory validation reports, and simulated-image feedback to support autonomous agent revisions. Its LLM4MR comparison denotes a configuration supplied with domain context but without tools~\cite{zaiss2026agent4mr}. Agent4MR also reports autonomous research experiments in which an agent modifies sequences and reconstruction under time constraints, combines grid search with MRzero differentiable optimization, and evaluates improvements using target-image error. It therefore covers automated development with physical feedback and objective-based optimization and provides an important basis for positioning our work.

SmartBloch-Agent differs from Agent4MR primarily in its internal physical-computation capability and the organization of the end-to-end research process. Our framework integrates the GPU-accelerated SmartBloch Full-Bloch engine developed in this work, allowing magnetization evolution and signal calculations for candidate sequences to run internally. Without involving external MRI simulation software in iterative loops, the agent coordinates sequence construction, physical computation, result analysis, design revision, or parameter search according to the task. Structural design and parameter optimization share this internal physical-computation capability, while event execution, signal acquisition, reconstruction, and scientific evaluation are configured for the sequence type and objective. The framework brings executable sequence design files and their simulation, reconstruction, and evaluation results into an integrated research workflow. Table~\ref{tab:related-comparison} compares representative approaches.

\begin{table}[!htbp]
\centering\footnotesize
\caption{Tasks and automation approaches in representative studies, compared according to their stated research objectives, computational pathways, and feedback mechanisms.}
\label{tab:related-comparison}
\setlength{\tabcolsep}{3pt}\renewcommand{\arraystretch}{1.16}
\begin{tabularx}{\linewidth}{@{}p{2.65cm}YYY@{}}
\toprule
Representative work & Main task & Physical execution and feedback & Decision and search approach\\
\midrule
Pulseq (2017)~\cite{layton2017pulseq} & Event representation and sequence prototyping & Scanner-side interpretation and execution & Researcher-led design and programming\\
MRzero (2021)~\cite{loktyushin2021mrzero} & Joint sequence and reconstruction design & Differentiable Bloch model and experimental validation & Gradient-based optimization\\
MR-double-zero (2022)~\cite{glang2022mrdoublezero} & Target contrast search & Measured scanner feedback & Automated derivative-free search\\
KomaMRI (2023)~\cite{castillopassi2023komamri} & General MRI numerical simulation & CPU/GPU physical solvers and reconstruction & Simulation platform\\
GPT4MR (2024)~\cite{zaiss2024gpt4mr} & Sequence and reconstruction code assistance & Domain prompting and interactive checks & Human--AI collaborative programming\\
SeqGPT (2025)~\cite{hussain2025seqgpt} & Sequence-structure generation (conference abstract) & Reported simulation of generated sequences & Generative model\\
Agent4MR (2026)~\cite{zaiss2026agent4mr} & Sequence development and objective-driven refinement (preprint) & Structured validation and image feedback & Agent-coordinated tools and iteration\\
SmartBloch-Agent & Automated sequence design and optimization & Integrated GPU-accelerated Full-Bloch engine; task-specific execution and evaluation & Agent-orchestrated design, optimization, and feedback decisions\\
\bottomrule
\end{tabularx}
\end{table}
\FloatBarrier

\section{The SmartBloch-Agent Framework}\label{sec:method}
\subsection{Framework Overview}\label{sec:overview}
\begin{figure}[!htbp]
\centering
\includegraphics[width=\linewidth]{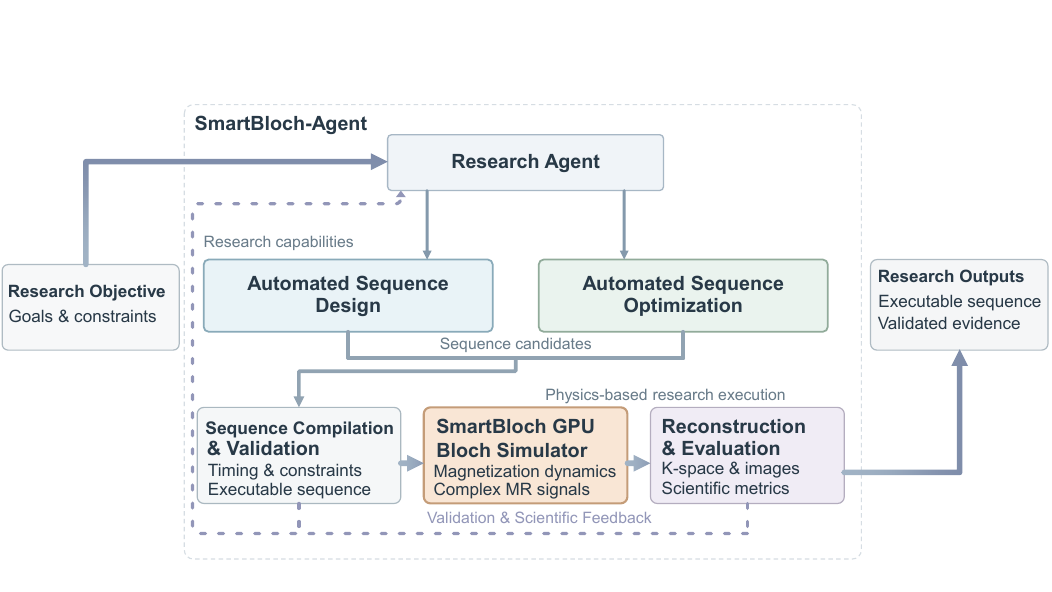}
\caption{Integrated SmartBloch-Agent workflow for MRI sequence research. The research agent interprets MRI objectives and coordinates sequence design and numerical optimization. The integrated GPU-accelerated SmartBloch Full-Bloch engine supports physical computation. Event execution, acquisition, reconstruction, and scientific evaluation are configured for each sequence and objective; their results guide subsequent agent decisions. Final outputs can be assembled into sequence packages for real-scanner workflows, together with supporting physical evidence.}
\label{fig:framework}
\end{figure}

SmartBloch-Agent is an integrated framework for automated MRI sequence research, connecting natural-language research intentions to executable sequences, physical simulation, reconstruction-based evaluation, and final outputs. As shown in Fig.~\ref{fig:framework}, the agent interprets requirements, plans tasks, orchestrates tools, and chooses subsequent actions from feedback. A sequence construction module converts designs into executable events; a numerical optimization module searches specified design spaces; and physical computation and evaluation modules turn candidates into signals, images, and quantitative feedback. Task states and research artifacts connect these modules, reducing manual coordination during tool invocation, result transfer, and workflow handoffs.

The framework brings two related activities into a common process. \emph{Sequence design} constructs and revises event composition, timing relationships, and acquisition strategies, whereas \emph{sequence optimization} searches variables under performance objectives and constraints. These activities can address separate tasks or proceed together through design, evaluation, and refinement. For different objectives, task specifications, sequence representations, available tools, and evaluation criteria configure the research process, allowing physical evidence to inform distinct types of research actions.

To describe research tasks consistently, we define the task contract in Eq.~\eqref{eq:contract}:
\begin{equation}
 \mathcal T=(\mathcal O,\mathcal Q,\mathcal C,\mathcal S,\mathcal M,\mathcal P),
 \label{eq:contract}
\end{equation}
Here, $\mathcal O$ denotes scientific objectives; $\mathcal Q$ denotes candidate-sequence representations and the permitted design space; $\mathcal C$ denotes physical, hardware, and timing constraints; $\mathcal S$ denotes simulation and evaluation scenarios; $\mathcal M$ denotes metrics and optimization directions; and $\mathcal P$ denotes eligibility, comparison, and selection criteria. This contract accommodates event-structure construction, waveform and parameter modification, and multiobjective evaluation. The agent can select research actions flexibly in response to task state, while executable checks and physical evaluations determine whether candidates satisfy scientific requirements.

\subsection{Agent-Driven Research Workflow}\label{sec:workflow}
The agent translates natural-language objectives into executable actions and updates its research strategy from task state, existing candidates, and evaluation feedback. In sequence design, actions include selecting or combining sequence components, modifying event dependencies, adjusting acquisition strategies, and correcting designs that violate constraints. In optimization, the agent orchestrates numerical search tools, interprets multiobjective results, and selects directions for further exploration. Its central role is research-level planning and cross-tool orchestration; specialized modules handle sequence construction, numerical search, Bloch evolution, reconstruction, and metric calculation. The agent is implemented with DeepSeek V4 Flash (\texttt{deepseek-v4-flash}) through an OpenAI-compatible function-calling interface with default API parameters; model configuration and record handling are described in Supplementary Section S7.

Let $q_t$ denote the candidate at iteration $t$. The task contract $\mathcal T$ determines the required validation, physical computation, acquisition, reconstruction, and metric extraction. A task-specific operation chain can be summarized as Eq.~\eqref{eq:physical-loop}:
\begin{equation}
 q_t\xrightarrow{\mathcal V_{\mathcal T}}\widehat q_t
 \xrightarrow{\mathcal B_{\mathcal T}}y_t
 \xrightarrow{\mathcal R_{\mathcal T}}z_t
 \xrightarrow{\mathcal E_{\mathcal T}}m_t .
 \label{eq:physical-loop}
\end{equation}
Here $\widehat q_t$ is an executable sequence that has passed the relevant checks, $y_t$ denotes the physical computation or acquisition output required by the task, $z_t$ denotes images or other reconstruction products when reconstruction is needed, and $m_t$ is the scientific metric vector defined by the research objective. This expression represents the logical workflow, while the actual steps vary by sequence and objective. When validation fails, structural conflicts, timing violations, and constraint reports also become feedback for candidate revisions.

Let $H_t$ denote the accumulated research history, including executed actions, candidate representations, validation outcomes, physical artifacts, and scientific metrics. Based on $(\mathcal T,H_t)$, the agent selects its next research action, calls a design tool or numerical optimizer, and incorporates new evidence into later decisions. A numerical optimizer can explore a parameter space continuously, while the agent coordinates tools when task planning, interpretation of feedback, or design changes are needed. This creates a closed loop of proposing a design, executing computations, interpreting results, and continuing exploration.

Research orchestration and scientific adjudication have distinct responsibilities. The agent can interpret feedback, invoke tools, and propose candidates, but its proposals cannot directly overwrite established physical results or evaluation rules. Evaluation modules enforce the scenarios, metric directions, and eligibility conditions in task contract\,$\mathcal T$\,, so candidates from different sources are assessed according to the task's physical computations and selection criteria. Additional archived hybrid-search experiments record actual structured tool actions, rejected invalid parameters, and failed scientific eligibility checks; examples and record units are provided in Supplementary Sections S3 and S4.

\subsection{Automated Sequence Design}\label{sec:design}
Automated sequence design converts research requirements into executable pulse sequences that satisfy physical and acquisition constraints. The framework first interprets imaging objectives, desired signal or contrast characteristics, acquisition conditions, and resource requirements from natural language and organizes them into a structured design task. This task specifies both the intended scientific outcome and which parts of the sequence may be adjusted, together with the constraints they must satisfy, thereby linking research objectives to event construction and evaluation.

The framework represents sequences through both structure and parameters, expressing a design as Eq.~\eqref{eq:sequence-compile}:
\begin{equation}
 q=(h,\theta),\qquad s=\operatorname{Compile}(h,\theta),
 \label{eq:sequence-compile}
\end{equation}
Here $h$ represents sequence-event composition and organization, including excitation, preparation, gradient encoding, readout, repeated structures, and temporal dependencies. $\theta$ contains design variables associated with these events, such as flip angles, durations, gradient waveforms, sampling, and timing. $s$ is the compiled executable sequence. Structural design combines sequence components, changes event order and dependencies, configures acquisition layouts, and modifies event organization and acquisition structure through new event combinations.

From the structured task, the agent selects design operations and proposes candidates. Sequence construction tools translate each candidate into explicit RF, gradient, sampling, and delay events and then check event relationships, timing consistency, hardware bounds, and acquisition layouts. The resulting event representation supports physical simulation and establishes a direct link between the executable sequence and subsequent research results. If structural, timing, or physical constraints are violated, the feedback identifies the affected events and reasons; the agent then coordinates revisions and repeats the checks.

Candidates that pass executability checks proceed to physical simulation and reconstruction. Scientific evaluation converts Bloch-derived signal evolution, acquisition data, and reconstructed images into contrast, signal-response, spatial-encoding, acquisition-efficiency, or other task-specific metrics. The agent uses these metrics and the research requirements to select subsequent design actions, such as revising event structures, changing sampling strategies, reconfiguring waveforms and timing, or optimizing parameters within an established structure. Physical feedback thus drives iterative sequence revision.

Design outputs comprise executable sequences, event and acquisition descriptions, key design parameters, physical simulation results, reconstruction products, and scientific evaluations. These artifacts can be assembled into a sequence output package for a real-scanner workflow. Changes to individual candidates remain linked to their evaluations, providing both executable designs and evidence of their physical behavior and selection. Structural design and parameter refinement can alternate within a task, allowing the research process to proceed from intention to verifiable designs and deliverables without manually coordinating intermediate tool handoffs.

\subsection{Automated Sequence Optimization}\label{sec:optimization}
Sequence optimization searches for better-performing protocols within a defined scientific objective and candidate space while keeping research requirements consistent. Variables may include continuous parameters of an existing design, discrete acquisition choices, waveform controls, or combinations thereof. The framework connects these variables and their valid ranges to executable sequence construction so that each candidate can be converted into event sequences and physically simulated using SmartBloch. Subsequent acquisition, reconstruction, and metric evaluation are configured for the research task.

The agent and numerical optimizers have complementary roles. Numerical algorithms search systematically within a specified variable space, whereas the agent uses interim metrics, trade-offs among candidates, constraints, and historical feedback to adjust subsequent actions. Multiobjective searches may use Optuna and NSGA-II to generate candidate sets~\cite{akiba2019optuna,deb2002nsgaii}; other strategies can be selected according to the search space and evaluation cost. Whether proposed by a language model, numerical search, or design revision, candidates undergo consistent executability checks and scientific evaluation to yield comparable outcomes. For flip-angle trajectories defined by a few control points, parameterization and interpolation expand low-dimensional candidates into complete waveforms across acquisition phases; constraints can be placed on changes between adjacent control points and adjacent phase-encoding steps separately.

The system generates search feedback from metric directions, candidate feasibility, and multiobjective trade-offs and compares and selects candidates satisfying scientific constraints. Physical evaluation and multiobjective feedback connect parameter optimization to sequence design and support iterative refinement.

\subsection{Physics-Based Evaluation and Feedback}\label{sec:evaluation}
Physical evaluation connects sequence protocols to imaging outcomes. Executable sequences specify RF excitation, gradient encoding, delays, and sampling events. Bloch simulation calculates magnetization evolution and signal response under specified tissue and field conditions, and the acquired data are reconstructed into images, signal curves, or other research products. Evaluation converts design and optimization candidates into quantitative physical observations that can be compared and used in subsequent decisions.

Scientific evaluation is configured by the task objective and can cover structural and timing validity, signal or tissue contrast, image quality, parameter sensitivity, field robustness, scan time, and RF energy. The workflow first checks executability, then computes task-specific physical metrics and applies scientific eligibility criteria. Eligible candidates proceed to multiobjective comparison and selection. Let $\mathcal Q_{\rm obs}$ denote the set of evaluated candidates, $V(q)$ indicate execution validity, and $G(m(q))$ indicate scientific eligibility. The eligible set is defined by Eq.~\eqref{eq:eligible}:
\begin{equation}
 \mathcal E_{\rm obs}=\bigl\{q\in\mathcal Q_{\rm obs}\,\bigm|\,V(q)=1,\,G(m(q))=1\bigr\}.
 \label{eq:eligible}
\end{equation}
For tasks with an aggregate preference function $S(q)$, the final selection is given by Eq.~\eqref{eq:selection}:
\begin{equation}
 q^{\star}=\underset{q\in\mathcal E_{\rm obs}}{\operatorname{arg\,max}}\;S(q).
 \label{eq:selection}
\end{equation}
This procedure separates \emph{executability}, \emph{scientific eligibility}, and \emph{preference ranking}. Compilation and physical execution establish validity, task constraints define the acceptable set, and selection is performed only among eligible candidates. Where objectives compete, Pareto relations can characterize trade-offs within the eligible set before a final selection is made according to the research objective. A selected sequence may also be executed separately to check consistency across events, physical outputs, and scientific metrics. Metrics, eligibility rules, and evaluation procedures are configured for each sequence and research task.

Quantitative optimization can also assess local signal sensitivity to parameters. Let $\boldsymbol{u}(\boldsymbol\theta)$ denote real-valued observed features extracted from simulated signals or reconstructions, and let $J=\partial\boldsymbol{u}/\partial\boldsymbol\theta$ denote the parameter Jacobian. For observation-noise covariance $\Sigma$, the Fisher information matrix and corresponding covariance bound are given by Eq.~\eqref{eq:fisher-information}:
\begin{equation}
 F=J^{\mathsf T}\Sigma^{-1}J,\qquad C=F^{-1}.
 \label{eq:fisher-information}
\end{equation}
The diagonal entry $C_{pp}$ or its square root describes local uncertainty in parameter $p$. Feature selection, noise assumptions, parameter scaling, and finite-difference step sizes determine the interpretation of this measure. Each optimization task uses normalized sensitivity metrics specified in its evaluation protocol. Feature construction, finite differences, and scientific eligibility criteria are detailed in Supplementary Section S2.

Evaluation feedback contains not just a final score but also signals, reconstructions, parameter changes, constraint checks, and comparisons among candidates. The agent uses this information to decide whether to continue design, start parameter search, revise the research protocol, or select an output. When needed, the selected sequence is recompiled, simulated, reconstructed, and evaluated under the recorded research conditions to check consistency. By linking research actions with physical artifacts, SmartBloch-Agent establishes a closed-loop process from candidate generation to scientific outcomes.

\subsection{The SmartBloch Simulation Engine}\label{sec:engine}
SmartBloch is a GPU-accelerated Bloch simulation engine developed in this study. It provides internal magnetization evolution and signal computation for candidate sequences. Event execution, sampling organization, reconstruction, and scientific evaluation are configured according to the sequence and research objective. The agent coordinates research operations and interprets feedback, while the sequence builder, numerical optimizer, and simulation engine handle structural generation, parameter search, and physical computation, respectively.

\textbf{Magnetization evolution and discrete-event computation.} SmartBloch represents an imaging object as a spatially distributed ensemble of spins. Each spin has an equilibrium magnetization $M_0$, longitudinal and transverse relaxation times $T_1$ and $T_2$, a spatial position, and local field properties. Its magnetization vector $\mathbf M=(M_x,M_y,M_z)^{\mathsf T}$ follows the Bloch equation in Eq.~\eqref{eq:smartbloch-equation}:
\begin{equation}
 \frac{\mathrm d\mathbf M}{\mathrm dt}
 =\gamma\,\mathbf M\times\mathbf B_{\mathrm{eff}}(t)
 -\begin{pmatrix}
  M_x/T_2\\[2pt]
  M_y/T_2\\[2pt]
  (M_z-M_0)/T_1
 \end{pmatrix},
 \label{eq:smartbloch-equation}
\end{equation}
Here $\gamma$ is the gyromagnetic ratio, and $\mathbf B_{\mathrm{eff}}$ incorporates RF excitation, position-dependent gradient contributions, and local main-field off-resonance. Sequence events are expanded into ordered discrete evolution steps. At each substep, magnetization orientation is updated by the Rodrigues rotation formula and longitudinal and transverse relaxation by closed-form exponential expressions. For finite-duration RF pulses, RF amplitude and phase, local $B_1$, gradients, and off-resonance jointly determine the effective rotation field. This implementation advances continuous magnetization dynamics using a piecewise-constant-field approximation; accuracy depends on event discretization and substep settings.

To represent intravoxel frequency distributions, the engine supports multiple weighted off-resonance components per spin and combines their complex signals during sampling. Deterministic quadrature at fixed off-resonance nodes approximates intravoxel dephasing contributions and provides repeatable component summation along a fixed computational path for given inputs.

\textbf{Event execution, signal sampling, and reconstruction interface.} The internal sequence builder and Pulseq import pathway organize RF excitation, free evolution, gradient encoding, and analog-to-digital converter (ADC) readouts into an event stream. The solver advances spatial magnetization along the event timeline, aggregates transverse magnetization into complex signals at sampling times, and performs demodulation according to receive phase and frequency. Samples are organized as complex $k$-space according to the sequence's spatial encoding and readout order, then passed to the reconstruction module to produce images and other artifacts needed for evaluation. Sequence events, complex samples, reconstructed images, and scientific metrics can therefore be linked within a single task record for interpretation and further agent decisions.

\textbf{GPU parallelism and runtime data organization.} SmartBloch uses CUDA to distribute independent spin trajectories across GPU threads. The conventional path maps one thread to one spin, with each thread advancing its magnetization state through the full event stream. In the diffusion path, a thread block is assigned to each spin and cooperatively updates subspins and aggregates their contributions within the block. The anatomical flow-coupled path processes events in batches and retains spin states within each batch to reduce state transfers at event boundaries. For anatomical flow, preloaded four-dimensional velocity fields provide local velocities through spatial and temporal interpolation; spin positions are advanced by the midpoint method, with precomputed vessel-segment indices supporting position projection. These pathways share sequence-event and signal representations but employ distinct thread layouts and update strategies for ordinary evolution, diffusion, and flow workloads.

At the beginning of each simulation call, the required spin, event, sampling, and field data are transferred to the device. Magnetization evolution and signal accumulation then occur on the GPU throughout event execution, and outputs are returned to the host at completion. The conventional path accumulates $k$-space signals through atomic operations. The anatomical flow path can alternatively store per-spin contributions and reduce them in a fixed order, yielding repeatable summation order at the cost of additional intermediate buffers. GPU performance reflects spin-level parallelism, analytical rotation and relaxation updates, continuous device-side execution, and specialized pathways for different physical workloads.

\textbf{Role in automated research.} SmartBloch receives compilation-validated candidate sequences and specified research conditions, performs the relevant Bloch computation, and returns data for downstream processing. Acquisition, reconstruction, and metric computation depend on the sequence and objective. The agent uses these outputs to decide whether to continue design, invoke parameter search, revise acquisition layouts, or select candidates; numerical optimizers perform searches in their designated variable spaces. Numerical consistency and GPU simulation efficiency are evaluated in Section~\ref{sec:results-kernel}.

\section{Results}\label{sec:results}
\subsection{Automated MRI Sequence Design}\label{sec:results-design}

\textbf{Research task and autonomous design.} We evaluated SmartBloch-Agent's ability to construct and revise executable sequences for imaging objectives using two-dimensional Cartesian fast spin echo (FSE). The task required a synthetic brain phantom to exhibit decreasing signal intensity from cerebrospinal fluid (CSF) to gray matter (GM) to white matter (WM), while satisfying scientific requirements including signal retention of at least $0.30$ and scan time no longer than $320$\,s. A fixed BrainWeb digital brain phantom and tissue regions of interest were used, with an acquisition matrix of $256\times256$. The agent proposed sequence structure and timing from the objective and then determined subsequent design actions using task-specific physical computation, acquisition, reconstruction, and scientific evaluation feedback.

\textbf{Design evolution driven by physical feedback.} Initial candidate A1 used a repetition time (TR) of $5$\,s and an echo spacing (ESP) of $20$\,ms, with the central $k$-space line acquired by the $4$th echo of shot~16 (shots indexed from $0$). In response to feedback, A2 reduced TR to $3$\,s and assigned the central phase-encoding line to the fifth echo of shot~0. A3 further increased ESP to $25$\,ms. The effective echo times ($TE_{\mathrm{eff}}$) were $80.004$, $100.004$, and $125.004$\,ms, respectively. All three candidates met the predefined scientific eligibility criteria, and A3 was selected for independent replay. Over this design sequence, normalized CSF/GM contrast increased from $0.33758$ to $0.44858$, GM/WM contrast increased from $0.06358$ to $0.11669$, and scan time decreased from $160$\,s to $96$\,s. The agent used this feedback to revise timing and acquisition layout.

\begin{figure}[!htbp]
\centering
\includegraphics[width=\linewidth]{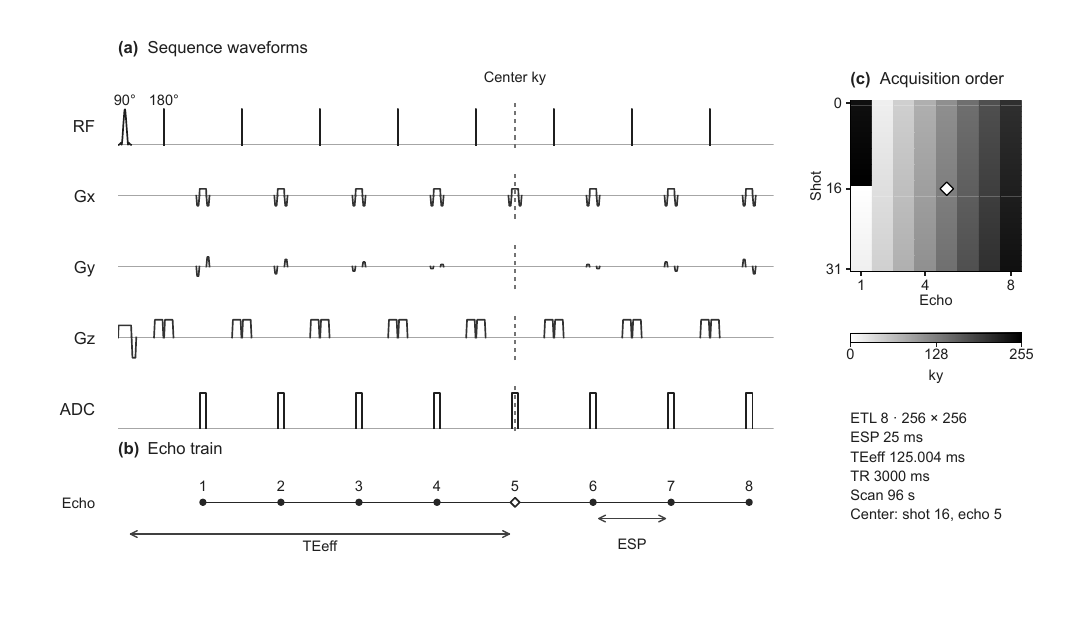}
\caption{Actual FSE sequence timing and acquisition layout. (a) RF, gradient, and sampling events; (b) the eight-echo train and effective echo time; and (c) banded phase-encoding order and the acquisition location of central $k$-space.}
\label{fig:fse-timing}
\end{figure}

\textbf{Executable timing and native acquisition.} The automatically designed candidates were executed as explicit RF, gradient, and ADC sampling events. Figure~\ref{fig:fse-timing} presents the actual FSE waveforms and acquisition layout for a banded phase-encoding variant of selected design A3. The sequence used $90^\circ$ excitation and $180^\circ$ refocusing pulses, an echo train length (ETL) of $8$, ESP of $25$\,ms, and TR of $3000$\,ms. The center of $k$-space was acquired at the $5$th echo of shot~16, with $TE_{\mathrm{eff}}=125.004$\,ms. The full acquisition consisted of $32$ shots and lasted $96$\,s. Native physical computation produced $256$ acquired phase-encoding lines with $256$ complex readout samples per line, yielding a $256\times256$ complex $k$-space array and its reconstructed image. The timing and phase-encoding layout establish the correspondence between sequence events and acquired data.

\begin{figure}[!htbp]
\centering
\includegraphics[width=\linewidth]{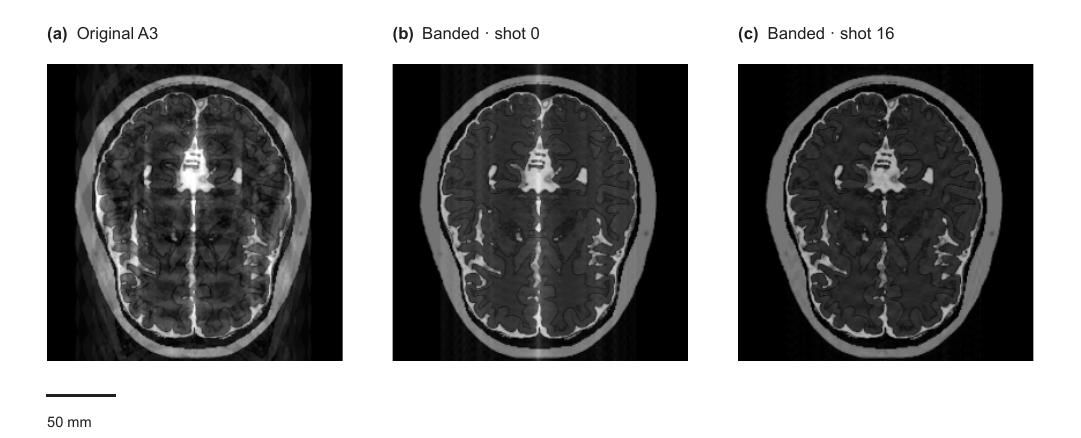}
\caption{Effect of phase-encoding layout on FSE reconstruction. (a) Original A3; (b) contiguous bands with the center in shot~0; and (c) contiguous bands with the center in shot~16. All images use the same display window.}
\label{fig:fse-ordering}
\end{figure}

\textbf{Effect of acquisition layout on image quality.} FSE echoes have complex signal responses that depend on their preceding magnetization history. The assignment of phase-encoding lines to echoes therefore directly affects the reconstructed image. We compared three phase-encoding orders for A3 under otherwise identical sequence and physical computation conditions. The original configuration used periodic echo assignment; two alternatives grouped the lines associated with each echo into contiguous bands and placed the central phase-encoding line in shot~0 or shot~16. Figure~\ref{fig:fse-ordering} compares their native reconstructions. The fraction of squared-magnitude energy outside the object support was calculated using Eq.~\eqref{eq:outside-energy}, where $I$ is the reconstructed image and $\Omega$ is the support formed by expanding the valid tissue region of the fixed phantom by four pixels:
\begin{equation}
R_{\mathrm{out}}=\frac{\displaystyle\sum_{(i,j)\notin\Omega}|I_{ij}|^2}{\displaystyle\sum_{i,j}|I_{ij}|^2}\times100\%.
\label{eq:outside-energy}
\end{equation}
The outside-support squared-magnitude energy fractions were $1.40180\%$, $0.24577\%$, and $0.04083\%$ for the three layouts. Relative to the original A3, contiguous banding with the center assigned to shot~16 reduced this energy by approximately $97.1\%$. Table~\ref{tab:fse-ordering-metrics} reports the outside-support energy and tissue contrast metrics. Improved spatial reconstruction was accompanied by redistributed tissue contrast: CSF/GM contrast changed from $0.44858$ to $0.50198$, whereas GM/WM contrast changed from $0.11669$ to $0.06021$.

\begin{table}[!htbp]
\centering\small
\caption{Outside-support energy and tissue contrast for different FSE phase-encoding layouts.}
\label{tab:fse-ordering-metrics}
\setlength{\tabcolsep}{8pt}\renewcommand{\arraystretch}{1.2}
\begin{tabular}{@{}lccc@{}}
\toprule
Phase-encoding layout & Outside energy (\%) & CSF/GM contrast & GM/WM contrast\\
\midrule
Original A3 & 1.40180 & 0.44858 & 0.11669\\
Contiguous bands, center at shot~0 & 0.24577 & 0.45128 & 0.07250\\
Contiguous bands, center at shot~16 & 0.04083 & 0.50198 & 0.06021\\
\bottomrule
\end{tabular}
\end{table}

\begin{figure}[!htbp]
\centering
\includegraphics[width=\linewidth]{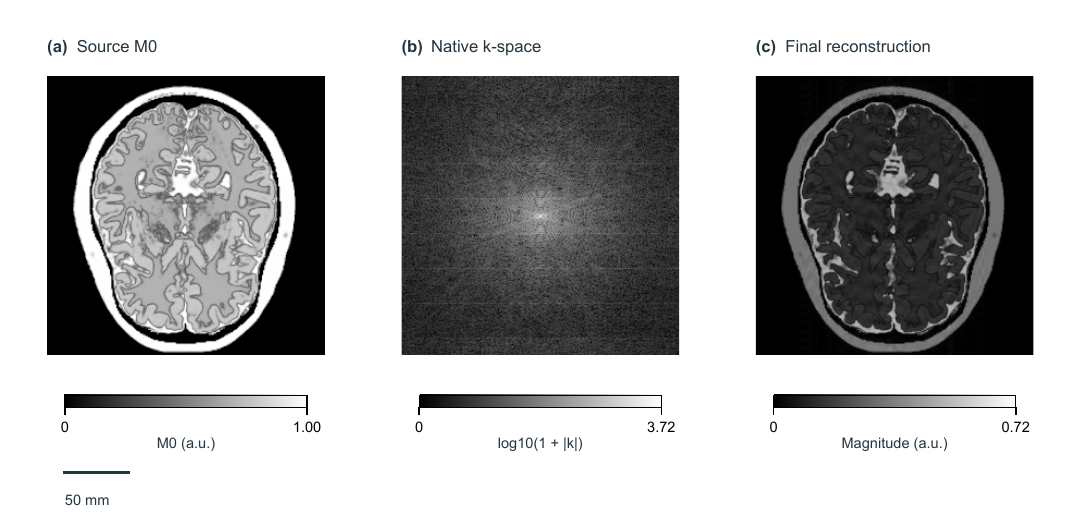}
\caption{Native acquisition and reconstruction for FSE sequence design. (a) Source equilibrium magnetization $M_0$; (b) logarithmic magnitude of native acquired $k$-space; and (c) final reconstructed magnitude image. Images and color scales were generated by the same simulation and reconstruction pipeline.}
\label{fig:fse-final-reconstruction}
\end{figure}

\textbf{Native acquisition and final reconstruction.} Figure~\ref{fig:fse-final-reconstruction} shows the source $M_0$ map, the magnitude display of native complex $k$-space, and the final reconstructed image, tracing the data path from sequence acquisition to image formation.

The original A3 configuration underwent native acquisition and independent replay on three fixed digital brain phantoms. Relative $L_2$ errors in the complex samples ranged from $1.90\times10^{-6}$ to $3.31\times10^{-6}$. This design process links agent proposals, revisions informed by physical feedback, executable sequences, and imaging evidence. The acquisition-layout comparison further shows how encoding order affects reconstruction.

\textbf{Repeated runs across design objectives.} Three independent design tasks were conducted for each of four predefined objectives: tissue contrast (FSE-A), scan time (FSE-B), RF resources (FSE-C), and spatial leakage (FSE-D). All $12$ planned tasks produced scientifically eligible designs that passed independent replay. Including interruptions caused by GPU resource occupancy, there were $15$ launch attempts. The first agent-generated candidate compiled successfully in $11/12$ tasks, and all three tasks requiring revision in response to scientific feedback ultimately produced eligible designs. No manual editing of scientific parameters occurred in completed tasks. Candidate counts, native evaluations, execution times, and resource recovery records are detailed in Supplementary Section S5.

In one feedback-based revision for a scan-time-constrained task, the agent adjusted echo train length and phase-encoding layout after image-error and tissue-signal-retention checks failed. Relative magnitude $L_2$ decreased from $0.18130$ to $0.08068$, and minimum tissue signal retention increased from $0.84351$ to $0.94530$. Scan time increased from $19.2$\,s to $38.4$\,s but remained below the task limit of $40$\,s. This revision shows how the agent modified an acquisition protocol in response to image-quality feedback while respecting the scan-time constraint.

\textbf{Workflow efficiency.} Table~\ref{tab:fse-workflow-compare} compares the execution time and human coordination costs of three FSE design workflows. SmartBloch-Agent achieved a mean end-to-end time of $44.47$\,min per task across four design objectives, without manual scientific-parameter editing. Under the corresponding task-specific coordination budgets, the fixed and manually coordinated workflows required approximately $80$ and $165$\,min per task, respectively, with approximately $25$ and $100$\,min of active human effort. The comparison highlights the contribution of automated sequence construction, feedback-driven revision, and scientific evaluation to research workflow efficiency.

\begin{table}[!htbp]
\centering\small
\caption{Workflow efficiency in FSE sequence design.}
\label{tab:fse-workflow-compare}
\setlength{\tabcolsep}{5pt}\renewcommand{\arraystretch}{1.2}
\begin{tabular}{@{}lrr@{}}
\toprule
Workflow & Human effort (min/task) & End-to-end time (min/task)\\
\midrule
SmartBloch-Agent & $0$ & $44.47$\\
Fixed workflow & $\approx25$ & $\approx80$\\
Manual coordination & $\approx100$ & $\approx165$\\
\bottomrule
\end{tabular}
\par\smallskip
\begin{minipage}{0.98\linewidth}\footnotesize
Human effort covers scientific-parameter adjustment, feedback interpretation, and research coordination. End-to-end time includes sequence construction, physical simulation, reconstruction, scientific evaluation, and workflow coordination. Agent runtime is recorded directly; the alternative workflows use task-specific coordination budgets detailed in Supplementary Section S8.1. Zero human effort denotes the absence of manual scientific-parameter editing and research decisions.
\end{minipage}
\end{table}

\subsection{Automated MRI Sequence Optimization}\label{sec:results-opt}

\textbf{Study system and sequential objectives.} We optimized a phase-encoding-dependent flip-angle trajectory for a five-echo steady-state free precession-based multiple overlapping-echo detachment (SSFP-MOLED) encoding mechanism under two-dimensional numerical acquisition~\cite{yang2026ssfpmoled}. Timing was fixed at $TR=18$\,ms and $TE=(3,6,9,12,15)$\,ms, with reference trajectory $\alpha_{\mathrm{ref}}(p)=15^\circ+0.06^\circ p$ ($p=0,\ldots,255$). Because the five-echo encoding exhibits different sensitivities to individual quantitative parameters, the study proceeded in two stages: first, evaluate the combined trade-offs among six parameters; then focus on $T_2$ and test whether improved sensitivity translates into downstream quantitative reconstruction. The agent organized research decisions and tool calls, while Optuna/NSGA-II searched the specified parameter space. All candidates underwent Full-Bloch simulation, complex reconstruction, and scientific evaluation, producing a shared archive for both stages.

\textbf{Stage 1: Combined six-parameter trade-off.} Of $191$ fully evaluated, deduplicated candidates, $58$ met the predefined scientific eligibility criteria. Under multiobjective preference ranking within this eligible set, language-model proposal \#353 was selected as the overall candidate. Relative to the fixed reference, the six normalized CRLB-related standard-deviation proxies changed as follows: $M_0$ increased by $0.0849\%$, while $T_1$, $T_2$, $T_2^*$, $B_0$, and $B_1$ decreased by $5.42\%$, $9.43\%$, $5.78\%$, $7.03\%$, and $19.31\%$, respectively. The candidate improved five of six metrics and achieved an aggregate score of $92.035636$ under the fixed scientific eligibility rules. Higher-scoring candidates failed signal, echo-balance, or robustness requirements and were excluded from the eligible set. Candidate screening and the full $17$-scenario replay are detailed in Supplementary Sections S2--S3.

\begin{figure}[!htbp]
\centering
\includegraphics[width=\linewidth]{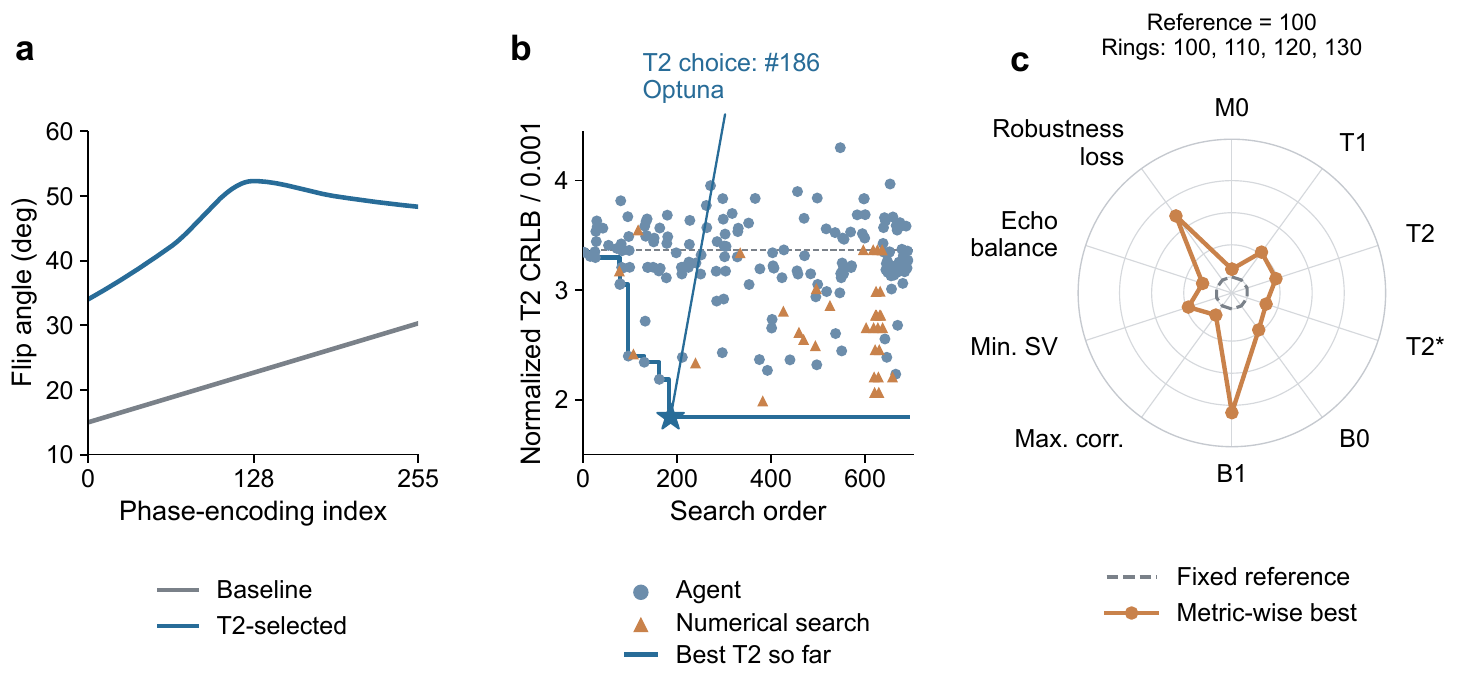}
\caption{Multiobjective and $T_2$-specific selection from the same candidate archive. (a) Flip-angle trajectories of the reference and $T_2$-specific candidate \#186; (b) trajectory of the $T_2$ standard-deviation-bound proxy for completed candidates in the original search archive; and (c) per-metric best envelope within the multiobjective eligible set, normalized to $100$ for the reference.}
\label{fig:opt-selection}
\end{figure}

\textbf{Stage 2: $T_2$-specific selection.} Following the combined trade-off analysis, candidates in the same archive that passed absolute execution checks and had complete $T_2$ evaluations were ranked specifically for the $T_2$ estimation objective. Search record \#186, corresponding to Optuna trial 244, reduced the normalized $T_2$ standard-deviation-bound proxy from $0.00336915$ for the reference to $0.00184196$, an improvement of $45.33\%$. This candidate was used for subsequent $T_2$ imaging and reconstruction to examine whether improved parameter sensitivity leads to a downstream quantitative reconstruction benefit. Figure~\ref{fig:opt-selection}a compares the flip-angle trajectories, Fig.~\ref{fig:opt-selection}b presents the archived $T_2$ metric trajectory, and Fig.~\ref{fig:opt-selection}c summarizes the independently achieved per-metric best envelope among eligible multiobjective candidates.

\textbf{Paired quantitative reconstruction evaluation.} To assess quantitative imaging with the $T_2$-specific candidate, we generated $512$ paired native Full-Bloch synthetic samples under each of the reference and optimized sequences, using matched sample identities and the same $384/64/64$ training, validation, and test split. Complex five-echo data were separated into ten real and imaginary input channels. Separate reconstruction mappings for $T_2$ were trained using the same R2AttUNet architecture, training budget, and model-selection rule. Both sequences were evaluated on the same $64$ held-out samples, with parameter error and image similarity calculated within fixed valid tissue masks. Metrics were computed for each sample and then averaged with equal weight.

\begin{figure}[!htbp]
\centering
\includegraphics[width=\linewidth]{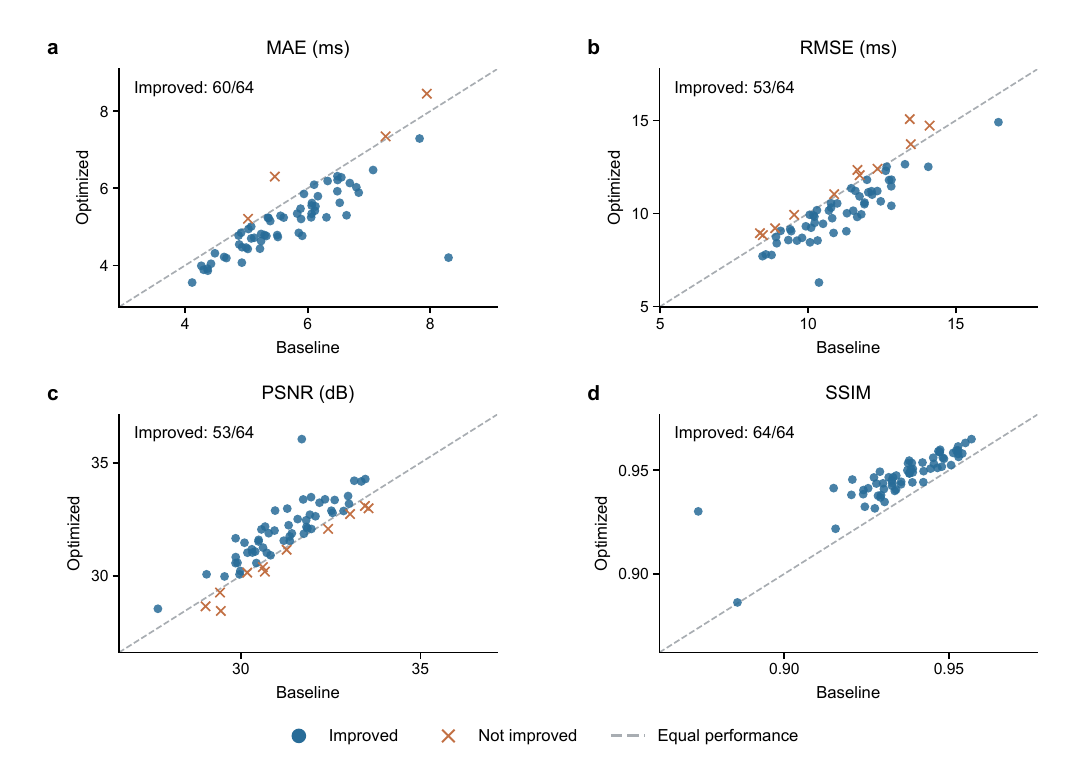}
\caption{Paired comparison of $T_2$ reconstruction metrics for reference and optimized sequences. The dashed line denotes equality; marker types distinguish improved and non-improved samples.}
\label{fig:opt-paired}
\end{figure}

Table~\ref{tab:opt-t2-test} shows that mean absolute error (MAE) decreased from $5.690$\,ms to $5.196$\,ms, an $8.69\%$ reduction. Root mean square error (RMSE), peak signal-to-noise ratio (PSNR), masked structural similarity (SSIM), and Pearson correlation also improved. Figure~\ref{fig:opt-paired} shows paired metrics across all $64$ test samples, with improvement in most cases.

\begin{table}[!htbp]
\centering
\small
\caption{Comparison of quantitative $T_2$ reconstruction for the reference and optimized sequences ($64$ paired held-out test samples).}
\label{tab:opt-t2-test}
\setlength{\tabcolsep}{9pt}\renewcommand{\arraystretch}{1.32}
\begin{tabular}{@{}lcccc@{}}
\toprule
Metric & Reference sequence & Optimized sequence & Mean change & Improved samples\\
\midrule
MAE (ms)$\downarrow$ & 5.690 & 5.196 & $-8.69\%$ & 60/64\\
RMSE (ms)$\downarrow$ & 11.097 & 10.368 & $-6.57\%$ & 53/64\\
PSNR (dB)$\uparrow$ & 31.201 & 31.820 & $+0.619$\,dB & 53/64\\
Masked SSIM$\uparrow$ & 0.9361 & 0.9471 & $+0.0110$ & 64/64\\
Pearson $r\uparrow$ & 0.9732 & 0.9778 & $+0.0046$ & 55/64\\
\bottomrule
\end{tabular}
\end{table}

The mean paired MAE difference was $0.4945$\,ms, with a median of $0.4543$\,ms; MAE increased in four cases. Leave-one-sample-out analysis gave a mean improvement range of $0.4373$--$0.5157$\,ms. The full paired distribution on synthetic test samples is presented in Supplementary Section S5.

\begin{figure}[!htbp]
\centering
\includegraphics[width=\linewidth]{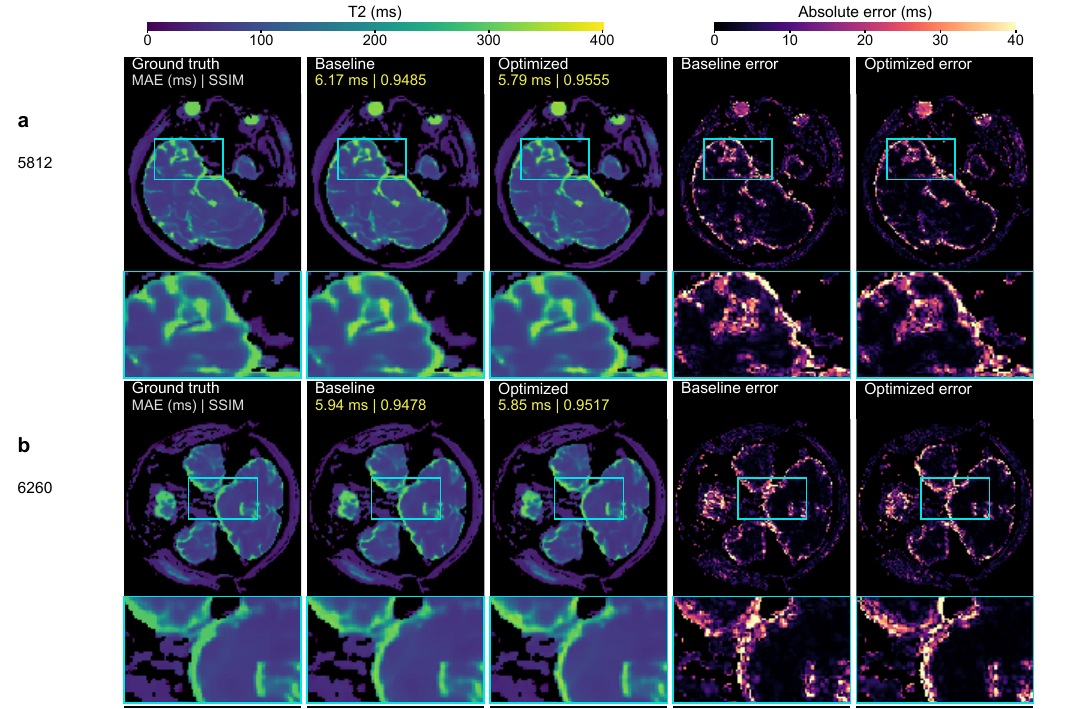}
\caption{Quantitative $T_2$ reconstruction and error distributions in two representative test samples. Identical color scales are used across columns; MAE and SSIM are calculated within the valid tissue region of each sample.}
\label{fig:opt-t2}
\end{figure}

\textbf{Quantitative image results.} Figure~\ref{fig:opt-t2} presents two test samples, showing $T_2$ ground truth, reconstructions with reference and optimized sequences, and corresponding absolute errors and local enlargements. Identical $T_2$ and error color scales are used to show tissue structure and spatial error distribution. MAE for the two samples decreased from $6.17$\,ms to $5.79$\,ms and from $5.94$\,ms to $5.85$\,ms, respectively. The error maps and enlarged views show how errors are distributed over tissue regions.

Under identical acquisition timing and matched reconstruction training rules, the selected flip-angle design reduced average $T_2$ reconstruction error, indicating that selection based on physical metrics can yield benefits for a downstream task. Its sum of squared flip angles was also approximately $4.124$ times that of the reference, reflecting an increase in the RF energy proxy alongside improved encoding performance. Quantitative accuracy and acquisition resources thus form a trade-off in sequence optimization.

\textbf{Cross-sequence workflow examples.} The same research orchestration was applied to parameter optimization of gradient echo (GRE) and spin echo (SE) sequences. GRE varied flip angle, TR, and TE, while SE varied TR and TE; both evaluated tissue contrast jointly with total acquisition time. Among six archived numerical-search/agent-assisted configurations for each family, five GRE configurations were completed and one was rejected for timing constraints; all six SE configurations were completed. Independent replay passed the relevant computational checks for all $11$ completed tasks. These examples show how candidate construction, physical evaluation, and selection extend to different parameterized sequences. Protocols and per-configuration results are provided in Supplementary Section S4.

\textbf{Research coordination efficiency.} The SSFP-MOLED optimization involved $1{,}135$ evaluation scenarios across three fidelity stages. High-fidelity physical evaluation dominated the computational workload, with the F2 stage accounting for $296.63$\,h of the total recorded execution time of $298.71$\,h. To characterize research-level efficiency, coordination costs were considered separately from physical computation. As summarized in Table~\ref{tab:ssfp-workflow-compare}, SmartBloch-Agent autonomously coordinated candidate generation, scientific evaluation, feedback interpretation, and final selection without manual scientific-parameter editing. The coordination budgets for fixed and manually coordinated workflows were approximately $10$ and $36$\,h per study, respectively. The comparison highlights the contribution of autonomous research orchestration to reducing human involvement in computationally intensive, multiobjective MRI sequence optimization.

\begin{table}[!htbp]
\centering\small
\caption{Research coordination efficiency in SSFP-MOLED optimization.}
\label{tab:ssfp-workflow-compare}
\setlength{\tabcolsep}{5pt}\renewcommand{\arraystretch}{1.2}
\begin{tabularx}{\linewidth}{@{}p{3.2cm}p{3.7cm}X@{}}
\toprule
Workflow & Human effort (h/study) & Research coordination\\
\midrule
SmartBloch-Agent & $0$ & Autonomous\\
Fixed workflow & $\approx10$ & Predefined checkpoints\\
Manual coordination & $\approx36$ & Manual feedback-driven revision\\
\bottomrule
\end{tabularx}
\par\smallskip
\begin{minipage}{0.98\linewidth}\footnotesize
Human effort covers research configuration, scientific-feedback interpretation, strategy adjustment, and final candidate selection. The comparison uses a common physical-evaluation workload. SmartBloch-Agent coordinates these research activities autonomously without manual scientific-parameter editing. Detailed coordination budgets are provided in Supplementary Section S8.2.
\end{minipage}
\end{table}

\subsection{Numerical Validation and Computational Performance of SmartBloch}\label{sec:results-kernel}

\textbf{Numerical consistency of magnetization evolution.} We evaluated the SmartBloch CUDA engine against an independent double-precision affine Bloch matrix-exponential reference using $36$ classes of single-voxel tests under white matter, gray matter, and cerebrospinal fluid conditions, with $8$ complex echo values per class. Tests covered static off-resonance, $B_1$ variation, imperfect refocusing, and finite-duration RF excitation. Table~\ref{tab:kernel-accuracy} reports the maximum complex absolute errors normalized to $M_0=1$. Instantaneous- and finite-duration-RF cases reached $1.69\times 10^{-6}$ and $5.13\times 10^{-5}$, respectively, both within prespecified tolerance limits. Agreement with the independent reference supports the numerical use of Bloch computation to evaluate sequence candidates in this framework.

\begin{table}[H]
\centering\footnotesize
\caption{Single-voxel numerical comparison between the SmartBloch CUDA simulation engine and an independent double-precision Bloch reference.}
\label{tab:kernel-accuracy}
\setlength{\tabcolsep}{7pt}\renewcommand{\arraystretch}{1.16}
\begin{tabular}{@{}lccc@{}}
\toprule
RF condition & Cases (echoes) & Maximum complex absolute error & Prespecified tolerance\\
\midrule
Instantaneous RF & 27 (216) & $1.69\times10^{-6}$ & $2\times10^{-5}$\\
Finite-duration RF & 9 (72) & $5.13\times10^{-5}$ & $2\times10^{-4}$\\
\bottomrule
\end{tabular}
\par\smallskip
\begin{minipage}{0.98\linewidth}\footnotesize
Error is defined as $\max|S_{\mathrm{CUDA}}-S_{\mathrm{ref}}|$ with $M_0=1$. Validation covers the listed static single-voxel, RF, and finite off-resonance conditions.
\end{minipage}
\end{table}

\textbf{GPU simulation performance.} Table~\ref{tab:smartbloch-benchmark} reports measured execution times for the current SmartBloch CUDA engine on GRE and GRE-EPI sequences, alongside comparison times for MRiLab and SPROM. Based on these timings, the CUDA implementation was $12.46$--$30.51$ times faster than MRiLab.

\begin{table}[H]
\centering\footnotesize
\caption{GRE and GRE-EPI runtime measurements for the current SmartBloch CUDA engine and comparative results (NVIDIA RTX 2080 Ti; seconds).}
\label{tab:smartbloch-benchmark}
\setlength{\tabcolsep}{5pt}\renewcommand{\arraystretch}{1.13}
\begin{tabular}{@{}lrrrr@{}}
\toprule
Sequence & MRiLab & SPROM & In-house CUDA implementation & Speedup vs. MRiLab\\
\midrule
GRE & 10.220 & 192.320 & 0.820 & $12.46\times$\\
Single-shot GRE-EPI & 11.442 & 29.812 & 0.466 & $24.55\times$\\
Multishot GRE-EPI & 15.620 & 32.410 & 0.512 & $30.51\times$\\
\bottomrule
\end{tabular}
\par\smallskip
\begin{minipage}{0.98\linewidth}\footnotesize
The table lists runtime measurements for the current SmartBloch CUDA engine and comparison times for MRiLab and SPROM; speedups were calculated from the reported times.
\end{minipage}
\end{table}

\FloatBarrier

\section{Discussion}\label{sec:discussion}
The two research tasks show that MRI sequence automation depends on organizing research intentions, sequence construction, parameter search, physical computation, reconstruction-based evaluation, and result packaging into a continuous process. SmartBloch-Agent assigns research-level planning and tool orchestration to the agent, while the sequence compiler, numerical optimizer, and SmartBloch engine perform structural construction, parameter search, and Bloch computation. Evaluation feedback informs subsequent decisions. The FSE workflow comparison highlights the role of automated coordination in reducing sequence-design effort. In the computationally intensive SSFP-MOLED study, SmartBloch-Agent integrates research planning, scientific-feedback interpretation, and candidate selection while high-fidelity physical simulation remains the dominant computational cost.

The FSE design study makes this connection concrete. Across successive revisions, the agent adjusted repetition time, echo spacing, and the acquisition location of the central phase-encoding line, progressing from the initial candidate to an executable design meeting the specified objectives. Scan time decreased from $160$\,s to $96$\,s, while both CSF/GM and GM/WM contrast increased. Physical feedback directly informed choices about event timing and acquisition organization. The subsequent phase-encoding comparison revealed another aspect of sequence design: magnetization history and signal modulation across echoes affect spatial reconstruction through the assignment of $k$-space lines. Replacing periodic interleaving with contiguous bands and changing the central-line acquisition location substantially reduced repeated structures in the reconstruction. The fraction of squared-magnitude energy outside tissue support decreased from $1.40180\%$ to $0.04083\%$. CSF/GM contrast increased, however, while GM/WM contrast decreased, indicating that improved spatial consistency and preservation of a particular tissue contrast are distinct objectives. This trade-off motivates joint consideration of signal contrast, acquisition layout, and reconstruction quality in sequence design.

The SSFP-MOLED study addressed two successive questions. First, a six-parameter trade-off was evaluated under fixed scientific eligibility requirements; the selected candidate \#353 improved five of six local standard-deviation-bound proxies. Next, candidate \#186 was selected from the same archive for its $T_2$-specific metric and assessed in downstream quantitative reconstruction using paired synthetic data and separately trained reconstruction networks. Its $T_2$ standard-deviation-bound proxy decreased by $45.33\%$, and mean MAE across $64$ test samples decreased by $8.69\%$. This reflects the combined performance of the acquisition encoding and its matched reconstructor. The sum of squared flip angles was $4.124$ times that of the reference, indicating a trade-off between quantitative accuracy and an RF energy proxy. Additional GRE and SE tasks illustrate how the same research orchestration accommodates other parameterized sequence problems.

Together, the experiments show that automatic sequence design and optimization address different levels of research variables but can be coordinated by an agent within a single framework. Design establishes measurement protocols at the level of events and acquisition structure; optimization searches for suitable operating conditions in a performance space. The SmartBloch Full-Bloch engine supplies internal physical computation for both, while execution, acquisition, reconstruction, and evaluation are configured by sequence and objective. We verified numerical consistency of the SmartBloch CUDA engine against an independent double-precision reference under the tested single-voxel conditions and reported runtime comparisons with other simulators for specified GRE-family tests. Current experiments primarily use two-dimensional numerical acquisitions and synthetic phantoms. Further assessments of simulation accuracy and throughput over broader sequences, complete imaging conditions, and controlled computational loads would help extend this automated research approach.

\section{Conclusion}\label{sec:conclusion}
We presented SmartBloch-Agent, an integrated framework that combines natural-language objective interpretation, research-level planning, sequence design, numerical optimization, Bloch simulation, reconstruction-based evaluation, and result packaging for MRI sequence research. The agent plans and orchestrates the research process, while numerical optimizers and the physical engine perform parameter search and computation, respectively. Task-specific execution and evaluation feed results back to the agent. In FSE design, the framework progressed from research objectives to executable event structures through proposals, feedback-driven revisions, and native acquisition, reducing manual workflow coordination. The SSFP-MOLED study first assessed six-parameter trade-offs and then selected a $T_2$-specific candidate, reducing mean absolute error in quantitative reconstruction by $8.69\%$ across $64$ paired synthetic test cases. Autonomous orchestration integrated research decisions and physical evaluation throughout the computationally intensive optimization process without manual scientific-parameter editing. Local numerical comparisons between SmartBloch and an independent double-precision Bloch reference satisfied prescribed tolerances. The framework places the research path from natural-language objectives to executable sequences, physical evidence, and deliverable outputs under unified orchestration.

\begingroup\small\bibliographystyle{unsrt}\bibliography{references}\endgroup
@article{bloch1946induction,
  author={Bloch, Felix}, title={Nuclear Induction}, journal={Physical Review},
  volume={70}, number={7--8}, pages={460--474}, year={1946}, doi={10.1103/PhysRev.70.460}
}

@article{lauterbur1973imaging,
  author={Lauterbur, Paul C.}, title={Image Formation by Induced Local Interactions: Examples Employing Nuclear Magnetic Resonance},
  journal={Nature}, volume={242}, number={5394}, pages={190--191}, year={1973}, doi={10.1038/242190a0}
}

@article{haase1986flash,
  author={Haase, A. and Frahm, J. and Matthaei, D. and H{\"a}nicke, W. and Merboldt, K. D.},
  title={{FLASH} imaging. {Rapid} {NMR} imaging using low flip-angle pulses},
  journal={Journal of Magnetic Resonance}, volume={67}, number={2}, pages={258--266}, year={1986},
  doi={10.1016/0022-2364(86)90433-6}
}

@article{pruessmann1999sense,
  author={Pruessmann, Klaas P. and Weiger, Markus and Scheidegger, Markus B. and Boesiger, Peter},
  title={{SENSE}: Sensitivity encoding for fast {MRI}},
  journal={Magnetic Resonance in Medicine}, volume={42}, number={5}, pages={952--962}, year={1999},
  doi={10.1002/(SICI)1522-2594(199911)42:5<952::AID-MRM16>3.0.CO;2-S}
}

@article{layton2017pulseq,
  author={Layton, Kelvin J. and Kroboth, Stefan and Jia, Feng and Littin, Sebastian and Yu, Huijun and Leupold, Jochen and Nielsen, Jon-Fredrik and St{\"o}cker, Tony and Zaitsev, Maxim},
  title={{Pulseq}: A rapid and hardware-independent pulse sequence prototyping framework},
  journal={Magnetic Resonance in Medicine}, volume={77}, number={4}, pages={1544--1552}, year={2017}, doi={10.1002/mrm.26235}
}

@article{stocker2010jemris,
  author={St{\"o}cker, Tony and Vahedipour, Kaveh and Pflugfelder, Daniel and Shah, N. J.},
  title={High-performance computing {MRI} simulations}, journal={Magnetic Resonance in Medicine},
  volume={64}, number={1}, pages={186--193}, year={2010}
}

@article{castillopassi2023komamri,
  author={Castillo-Passi, Carlos and Coronado, Ronal and Varela-Mattatall, Gabriel and Alberola-L{\'o}pez, Carlos and Botnar, Ren{\'e} and Irarr{\'a}zaval, Pablo},
  title={{KomaMRI.jl}: An open-source framework for general {MRI} simulations with {GPU} acceleration},
  journal={Magnetic Resonance in Medicine}, volume={90}, number={1}, pages={329--342}, year={2023}, doi={10.1002/mrm.29635}
}

@article{ma2013mrf,
  author={Ma, Dan and Gulani, Vikas and Seiberlich, Nicole and Liu, Kecheng and Sunshine, Jeffrey L. and Duerk, Jeffrey L. and Griswold, Mark A.},
  title={Magnetic resonance fingerprinting}, journal={Nature}, volume={495}, number={7440}, pages={187--192}, year={2013}
}

@article{zhao2019oedmrf,
  author={Zhao, Bo and Haldar, Justin P. and Liao, Congyu and Ma, Dan and Jiang, Yun and Griswold, Mark A. and Setsompop, Kawin and Wald, Lawrence L.},
  title={Optimal experiment design for magnetic resonance fingerprinting: {Cram{\'e}r--Rao} bound meets spin dynamics},
  journal={IEEE Transactions on Medical Imaging}, volume={38}, number={3}, pages={844--861}, year={2019}
}

@article{cohen2017algorithms,
  author={Cohen, Ouri and Rosen, Matthew S.}, title={Algorithm comparison for schedule optimization in {MR} fingerprinting},
  journal={Magnetic Resonance Imaging}, volume={41}, pages={15--21}, year={2017}
}

@article{scopecrafts2022bsplinemrf,
  author={Crafts, Evan Scope and Lu, Hengfa and Ye, Huihui and Wald, Lawrence L. and Zhao, Bo},
  title={An efficient approach to optimal experimental design for magnetic resonance fingerprinting with {B}-splines},
  journal={Magnetic Resonance in Medicine}, volume={88}, number={1}, pages={239--253}, year={2022}
}

@article{loktyushin2021mrzero,
  author={Loktyushin, Alexander and Herz, Kai and Dang, Hoai Nam and Glang, Felix and Deshmane, A. and Weinm{\"u}ller, Simon and Doerfler, Arnd and Sch{\"o}lkopf, Bernhard and Scheffler, Klaus and Zaiss, Moritz},
  title={{MRzero}---automated discovery of {MRI} sequences using supervised learning},
  journal={Magnetic Resonance in Medicine}, volume={86}, number={2}, pages={709--724}, year={2021}, doi={10.1002/mrm.28727}
}

@article{luo2021autodiff,
  author={Luo, Tianrui and Noll, Douglas C. and Fessler, Jeffrey A. and Nielsen, Jon-Fredrik},
  title={Joint design of {RF} and gradient waveforms via auto-differentiation for 3D tailored excitation in {MRI}},
  journal={IEEE Transactions on Medical Imaging}, volume={40}, number={12}, pages={3305--3314}, year={2021}
}

@article{endres2024pdg,
  author={Endres, Jonathan and Weinm{\"u}ller, Simon and Dang, Hoai Nam and Zaiss, Moritz},
  title={Phase distribution graphs for fast, differentiable, and spatially encoded {Bloch} simulations of arbitrary {MRI} sequences},
  journal={Magnetic Resonance in Medicine}, volume={92}, number={3}, pages={1189--1204}, year={2024}
}

@article{glang2022mrdoublezero,
  author={Glang, Felix and Mueller, Sebastian and Herz, Kai and Loktyushin, Alexander and Scheffler, Klaus and Zaiss, Moritz},
  title={{MR-double-zero}---proof-of-concept for a framework to autonomously discover {MRI} contrasts},
  journal={Journal of Magnetic Resonance}, volume={341}, pages={107237}, year={2022}, doi={10.1016/j.jmr.2022.107237}
}

@inproceedings{yao2023react,
  author={Yao, Shunyu and Zhao, Jeffrey and Yu, Dian and Du, Nan and Shafran, Izhak and Narasimhan, Karthik R. and Cao, Yuan},
  title={{ReAct}: Synergizing reasoning and acting in language models}, booktitle={International Conference on Learning Representations}, year={2023}
}

@article{boiko2023coscientist,
  author={Boiko, Daniil A. and MacKnight, Robert and Kline, Ben and Gomes, Gabe},
  title={Autonomous chemical research with large language models}, journal={Nature},
  volume={624}, number={7992}, pages={570--578}, year={2023}
}

@inproceedings{zaiss2024gpt4mr,
  author={Zaiss, Moritz and Rajput, Junaid R. and Dang, Hoai N. and Golkov, Vladimir and Cremers, Daniel and Knoll, Florian and Maier, Andreas},
  title={Exploring {GPT-4} as {MR} Sequence and Reconstruction Programming Assistant: {GPT4MR}},
  booktitle={Bildverarbeitung f{\"u}r die Medizin 2024}, pages={94--99}, publisher={Springer Vieweg}, year={2024}
}

@misc{zaiss2026agent4mr,
  author={Zaiss, Moritz and Aly, Amr and Endres, Jonathan and Dornstetter, Tobias and Weinm{\"u}ller, Simon and Maier, Andreas},
  title={Agentic {MR} sequence development: leveraging {LLMs} with {MR} skills for automatic physics-informed sequence development},
  howpublished={arXiv preprint arXiv:2604.13282}, year={2026}, eprint={2604.13282}, archivePrefix={arXiv}
}

@article{ravi2018pulseqgpi,
  author={Ravi, Keerthi Sravan and Potdar, Sneha and Poojar, Pavan and Reddy, Ashok Kumar and Kroboth, Stefan and Nielsen, Jon-Fredrik and Zaitsev, Maxim and Venkatesan, Ramesh and Geethanath, Sairam},
  title={{Pulseq}-Graphical Programming Interface: Open source visual environment for prototyping pulse sequences and integrated magnetic resonance imaging algorithm development},
  journal={Magnetic Resonance Imaging}, volume={52}, pages={9--15}, year={2018}
}

@article{ravi2019pypulseq,
  author={Ravi, Keerthi Sravan and Geethanath, Sairam and Vaughan, John Thomas},
  title={{PyPulseq}: A {Python} Package for {MRI} Pulse Sequence Design}, journal={Journal of Open Source Software},
  volume={4}, number={42}, pages={1725}, year={2019}, doi={10.21105/joss.01725}
}

@misc{hussain2025seqgpt,
  author={Hussain, Snawar and Huber, J{\"o}rn and G{\"u}nther, Matthias and Hoinkiss, Daniel Christopher},
  title={{SeqGPT}: Training a large language model to generate {MRI} pulse sequences},
  howpublished={Proceedings of the International Society for Magnetic Resonance in Medicine, conference abstract 3381},
  year={2025}
}

@article{sommer2017encoding,
  author={Sommer, K. and Amthor, T. and Doneva, M. and Koken, P. and Meineke, J. and B{\"o}rnert, P.},
  title={Towards predicting the encoding capability of {MR} fingerprinting sequences},
  journal={Magnetic Resonance Imaging}, volume={41}, pages={7--14}, year={2017}
}

@article{kara2019maperror,
  author={Kara, Danielle and Fan, Mingdong and Hamilton, Jesse and Griswold, Mark and Seiberlich, Nicole and Brown, Robert},
  title={Parameter map error due to normal noise and aliasing artifacts in {MR} fingerprinting},
  journal={Magnetic Resonance in Medicine}, volume={81}, number={5}, pages={3108--3123}, year={2019}
}

@article{perlman2022autocest,
  author={Perlman, Or and Zhu, Bo and Zaiss, Moritz and Rosen, Matthew S. and Farrar, Christian T.},
  title={An end-to-end {AI}-based framework for automated discovery of rapid {CEST/MT} {MRI} acquisition protocols and molecular parameter quantification ({AutoCEST})},
  journal={Magnetic Resonance in Medicine}, volume={87}, number={6}, pages={2792--2810}, year={2022}
}

@article{liu2017mrilab,
  author={Liu, Fang and Velikina, Julia V. and Block, Walter F. and Kijowski, Richard and Samsonov, Alexey A.},
  title={Fast Realistic {MRI} Simulations Based on Generalized Multi-Pool Exchange Tissue Model},
  journal={IEEE Transactions on Medical Imaging}, volume={36}, number={2}, pages={527--537}, year={2017}, doi={10.1109/TMI.2016.2620961}
}

@article{kose2017blochsolver,
  author={Kose, Ryoichi and Kose, Katsumi},
  title={{BlochSolver}: A {GPU}-optimized fast 3D {MRI} simulator for experimentally compatible pulse sequences},
  journal={Journal of Magnetic Resonance}, volume={281}, pages={51--65}, year={2017}, doi={10.1016/j.jmr.2017.05.007}
}

@article{nurdinova2025gpujemris,
  author={Nurdinova, Aizada and Ruschke, Stefan and Gestrich, Michael and Stelter, Jonathan and Karampinos, Dimitrios C.},
  title={{GPU}-accelerated {JEMRIS} for extensive {MRI} simulations},
  journal={Magnetic Resonance Materials in Physics, Biology and Medicine}, volume={38}, number={4}, pages={679--694}, year={2025}
}

@article{macleod2020selfdriving,
  author={MacLeod, B. P. and Parlane, F. G. L. and Morrissey, T. D. and H{\"a}se, F. and Roch, L. M. and Dettelbach, K. E. and others},
  title={Self-driving laboratory for accelerated discovery of thin-film materials}, journal={Science Advances},
  volume={6}, number={20}, pages={eaaz8867}, year={2020}
}

@article{macleod2022pareto,
  author={MacLeod, Benjamin P. and Parlane, Fraser G. L. and Rupnow, Connor C. and Dettelbach, Kevan E. and Elliott, Michael S. and Morrissey, Thomas D. and others},
  title={A self-driving laboratory advances the {Pareto} front for material properties},
  journal={Nature Communications}, volume={13}, number={1}, pages={995}, year={2022}
}

@article{lu2026aiscientist,
  author={Lu, Chris and Lu, Cong and Lange, Robert Tjarko and Yamada, Yutaro and Hu, Shengran and Foerster, Jakob and Ha, David and Clune, Jeff},
  title={Towards end-to-end automation of {AI} research}, journal={Nature},
  volume={651}, number={8107}, pages={914--919}, year={2026}, doi={10.1038/s41586-026-10265-5}
}

@inproceedings{akiba2019optuna,
  author={Akiba, Takuya and Sano, Shotaro and Yanase, Toshihiko and Ohta, Takeru and Koyama, Masanori},
  title={{Optuna}: A next-generation hyperparameter optimization framework},
  booktitle={Proceedings of the 25th ACM SIGKDD International Conference on Knowledge Discovery and Data Mining},
  pages={2623--2631}, publisher={ACM}, year={2019}
}

@article{deb2002nsgaii,
  author={Deb, Kalyanmoy and Pratap, Amrit and Agarwal, Sameer and Meyarivan, T.},
  title={A fast and elitist multiobjective genetic algorithm: {NSGA-II}},
  journal={IEEE Transactions on Evolutionary Computation}, volume={6}, number={2}, pages={182--197}, year={2002}
}

@article{yang2026ssfpmoled,
  author={Yang, Jingying and Zhu, Liuhong and Xiong, Kai and Bao, Jianfeng and Yang, Qinqin and Chen, Weikun and Kang, Taishan and Zhou, Jianjun and Lin, Jianzhong and Lin, Liangjie and Chen, Zhong and Cai, Shuhui and Cai, Congbo},
  title={Rapid multi-parametric quantitative {MRI} via deep learning-based synthetic-to-real reconstruction and 3D {SSFP-MOLED} imaging},
  journal={NeuroImage}, volume={335}, pages={121985}, year={2026}, doi={10.1016/j.neuroimage.2026.121985}
}
\end{document}

\maketitle
\noindent This supplement provides numerical definitions, candidate-search archives, repeated design experiments, and additional validation supporting the methods and results in the main paper. Three groups of research evidence are distinguished: (A) automated FSE sequence design and the subsequent acquisition-order comparison; (B) the original steady-state free precession-based multiple overlapping-echo detachment (SSFP-MOLED) multiobjective hybrid search and archive-wide reselection of eligible candidates (record \#353); and (C) subsequent $T_2$-specific analysis and paired reconstruction using the same search archive (record \#186). GRE/SE extension tasks have separate research settings. The two SSFP-MOLED candidates serve different selection criteria and subsequent evaluation purposes.

\section{Study Identities and Corresponding Evidence}
\begin{table}[htbp]\centering\footnotesize
\caption{Study instances and their principal evidence in this paper.}\label{tab:scope}
\setlength{\tabcolsep}{4pt}\renewcommand{\arraystretch}{1.12}
\begin{tabularx}{\linewidth}{@{}>{\raggedright\arraybackslash}p{2.45cm}>{\raggedright\arraybackslash}p{3.4cm}>{\raggedright\arraybackslash}p{3cm}Y@{}}\toprule
Instance & Research subject and final selection & Main evaluation & Evidence location\\\midrule
A: FSE design & Agent revisions A1--A3; additional banded-order comparison based on A3 & Timing, tissue contrast, outside-support energy, same-backend replay & Main Section 4.1; Supplement S5\\
B: Archived multiobjective optimization & SSFP-MOLED, LLM candidate \#353 & Six local uncertainty measures, scientific eligibility, 17-scenario replay & Main Section 4.2; Supplements S2--S3\\
C: $T_2$-specific study & Optuna candidate \#186 from the same SSFP-MOLED archive & $T_2$ sensitivity and 64 paired reconstructions & Main Section 4.2; Supplement S5\\
D: Extension tasks & GRE and SE with numerical-search/agent-assisted configurations & Contrast--time trade-offs and replay & Supplement S4\\
SmartBloch simulation engine & CUDA computation and independent double-precision reference & Single-voxel complex error and selected spatial integration tests & Main Section 4.3; Supplement S6\\\bottomrule
\end{tabularx}
\end{table}

\section{Computational Definitions for Archived Multiobjective Optimization}
\subsection{Observed Features, Finite Differences, and Fisher Information}
This section describes the actual archived evaluator used in the original SSFP-MOLED multiobjective search. Records \#186 and \#353 were both generated in this search and share the same normalized sensitivity evaluation but were subsequently selected for different purposes. For each of the five echoes, the original archive produced a complex image $I_e\in\mathbb C^{256\times256}$ and an orientation-corrected complex $k$-space array $K_e$. For region $R$, the complex mean is $\mu(X,R)=|R|^{-1}\sum_{(y,x)\in R}X_{yx}$. The central image region is $C=[64,192)^2$; the quadrants are $Q_{ab}=[128a,128(a+1))\times[128b,128(b+1))$, $a,b\in\{0,1\}$; and the central $k$-space region is $C_k=[112,144)^2$. Indices are zero-based with excluded right endpoints. Five complex image means and three complex $k$-space means are extracted for each echo:
\begin{equation}
 z_e^I=(\mu(I_e,C),\mu(I_e,Q_{00}),\mu(I_e,Q_{01}),\mu(I_e,Q_{10}),\mu(I_e,Q_{11})),
\end{equation}
\begin{equation}
 z_e^K=(\mu(K_e,C_k),\mu(K_e,Q_{00}),\mu(K_e,Q_{11})),\quad z_e=(z_e^I,z_e^K).
\end{equation}
Across five echoes, $z\in\mathbb C^{40}$. Interleaving its real and imaginary components gives $u\in\mathbb R^{80}$. The regions overlap spatially, so these features represent local sensitivities within the same synthetic model rather than independent tissue measurements.

Central finite differences are used for $\theta=(M_0,T_1,T_2,T_2^*,B_0,B_1)$:
\begin{equation}
 J_{:p}=\frac{u(\theta+h_pe_p)-u(\theta-h_pe_p)}{2h_p},\quad P=\operatorname{diag}(s_p).
\end{equation}
\begin{table}[tbp]\centering\footnotesize
\caption{Parameter scales and central finite-difference steps in the original archive.}\label{tab:scales}
\begin{tabular}{@{}lccc@{}}\toprule Parameter & Unit & Scale $s_p$ & Step $h_p$\\\midrule
$M_0$ & Relative magnetization & $1$ & $0.01$\\
$T_1$ & s & $1\,\mathrm{s}$ & $0.02\,\mathrm{s}$\\
$T_2$ & s & $0.1\,\mathrm{s}$ & $0.002\,\mathrm{s}$\\
$T_2^*$ & s & $0.05\,\mathrm{s}$ & $0.002\,\mathrm{s}$\\
$B_0$ & Hz & $100\,\mathrm{Hz}$ & $1\,\mathrm{Hz}$\\
$B_1$ & Dimensionless & $1$ & $0.01$\\\bottomrule\end{tabular}
\end{table}
Let $a_i=|z_i|$ be the feature magnitude under nominal conditions, and set $f=\max\{0.05\operatorname{median}(a_i:a_i>0),10^{-12}\}$. For each complex feature, both its real and imaginary parts are scaled by $\max(a_i,f)$, defining the $80\times80$ diagonal matrix $D$. The archived production evaluator uses
\begin{equation}
 A=\frac{D^{-1}JP}{\eta},\quad \eta=0.01,\quad F=A^{\mathsf T}A,
\end{equation}
\begin{equation}
 A=U\operatorname{diag}(\sigma_j)V^{\mathsf T},\quad C=V\operatorname{diag}(\sigma_j^{-2})V^{\mathsf T},\quad r_p=\sqrt{\max(C_{pp},0)}.
\end{equation}
For the local calculation with fixed $D$, the assumed covariance of the original real-valued features is $\eta^2D^2$. Additional diagnostics are minimum singular value $\sigma_{\min}$, condition ratio $\kappa=(\sigma_{\max}/\sigma_{\min})^2$, and maximum absolute parameter correlation. $r_p$ is a CRLB-related \emph{standard-deviation proxy} in normalized coordinates, not a variance CRLB in the original physical units or the MAE of a reconstructed $T_2$ image. Singular or nonfinite information matrices are excluded from eligible ranking. Additional $T_2^*$ dephasing in this archive is approximated by $60$ discrete off-resonance subspins, as defined in the original archived method.

\subsection{Multiobjective Preference and Scientific Safeguards}
For a minimized metric $x$ with tolerance scale $a$, the records use $d(x)=1/[1+(x/a)^2]$; for a maximized metric, they use $d(x)=1/[1+(a/x)^2]$. The six $r_p$ values are aggregated as
\begin{equation}
 d_{\mathrm{CRLB}}=0.6\min_p d_p+0.4\left(\prod_p d_p\right)^{1/6}.
\end{equation}
The aggregate preference score is
\begin{equation}
 S=100\exp\!\left(\frac{\sum_i w_i\ln d_i}{\sum_i w_i}\right),
\end{equation}
The six weights correspond to aggregate CRLB, maximum correlation, minimum singular value, condition ratio, echo balance, and worst-scenario robustness, respectively: $(0.40,0.15,0.075,0.075,0.15,0.15)$. Physical and scientific checks precede final ranking; aggregate scores from different task protocols are not compared directly.

\begin{table}[htbp]\centering\footnotesize
\caption{Predefined scientific eligibility and reference-relative degradation safeguards in the original archive. At least one primary metric must improve by $3\%$.}\label{tab:gates}
\begin{tabular}{@{}p{5.6cm}ccc@{}}\toprule
Metric & Direction & Absolute criterion & Allowed degradation\\\midrule
Six $r_p$ values & Lower is better & Each $\leq0.25$ & $5\%$\\
Condition ratio $\kappa$ & Lower is better & $\leq10^6$ & $20\%$\\
Minimum singular value & Higher is better & $\geq0.01$ & $5\%$\\
Maximum parameter correlation & Lower is better & $\leq0.90$ & $5\%$\\
Echo balance & Higher is better & $\geq0.05$ & $5\%$\\
Robustness loss & Lower is better & $\leq0.50$ & $5\%$\\
Nominal full-image signal & Higher is better & $\geq10^{-8}$ & $3\%$\\
Band/stripe energy fractions & Lower is better & No absolute criterion & $10\%$\\
Total duration & Lower is better & $\leq120\,\mathrm{s}$ & ---\\\bottomrule
\end{tabular}
\end{table}
These conditions define the frozen scientific eligibility policy of the original archive.

\section{Archived SSFP-MOLED Multiobjective Search, Candidate Selection, and Replay}
\subsection{Search Scale and Final Candidate}
The original archive used a five-control-point piecewise cubic Hermite interpolating polynomial (PCHIP) flip-angle representation over $256$ phase-encoding lines. The fixed linear reference was $\alpha(p)=15^\circ+0.06^\circ p$. Control points were constrained to $[5^\circ,60^\circ]$, changes between adjacent control points to at most $20^\circ$, and changes between adjacent interpolated phase-encoding steps to at most $0.25^\circ$. Timing was fixed at $TR=18\,\mathrm{ms}$ and $TE=(3,6,9,12,15)\,\mathrm{ms}$ for the five echoes. Evaluation covered one nominal scenario, four $B_0/B_1$ scenarios ($B_0=\pm50\,\mathrm{Hz}$, $B_1=0.90,1.10$), and twelve central finite-difference scenarios consisting of positive and negative perturbations to six tissue parameters, for $17$ scenarios in total.

\begin{figure}[!htbp]
\centering
\includegraphics[width=0.97\linewidth]{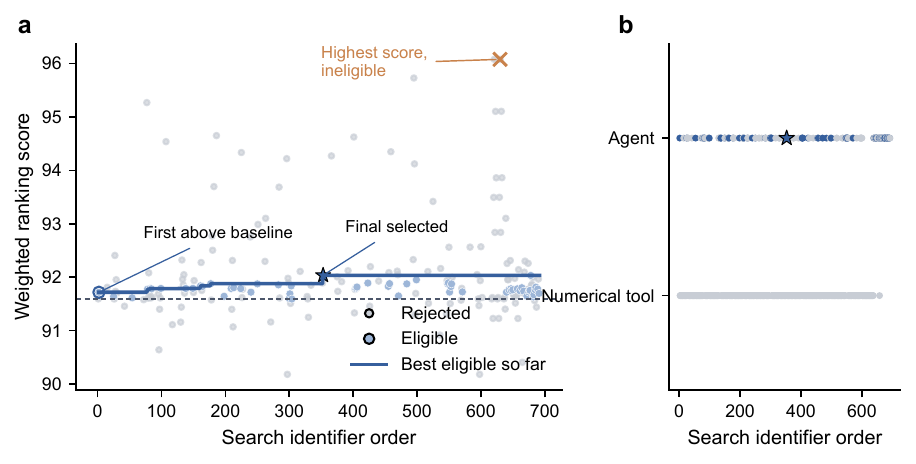}
\caption{Archived hybrid search and scientific eligibility in the original archive. (a) Aggregate scores, scientific eligibility states, and best-so-far scores within the eligible set, ordered by search identifier. The label ``highest score but ineligible'' denotes candidates violating scientific safeguards; ``final selection'' denotes the LLM proposal \#353 selected after archive review. (b) Candidate sources from the language model (agent) and numerical tools. Search identifiers include rejected and incomplete records and therefore do not equal the number of independently completed evaluated sequences.}
\label{fig:archive-search}
\end{figure}

The archive contained $691$ search identifiers, including $180$ LLM-generated and $511$ numerical-search records. Of these, $474$ were rejected before Bloch evaluation, $212$ reached a post-Bloch terminal state, and $5$ remained unfinished. Recovery and deduplication produced $194$ candidate sequences, of which $191$ had complete evaluations. A uniform archive review found $58$ candidates meeting the frozen scientific eligibility rules. The finalizer used in the original recovery stage and the subsequent archive-wide reranking were separate operations. The archived best candidate reported here was obtained from the latter. Figure~\ref{fig:archive-search} shows the search trajectory and candidate sources. Its horizontal axis represents search identifiers rather than sequentially numbered completed independent sequences; the marked final candidate reflects eligibility review and reranking of the complete archive.

\begin{figure}[!htbp]
\centering
\includegraphics[width=0.97\linewidth]{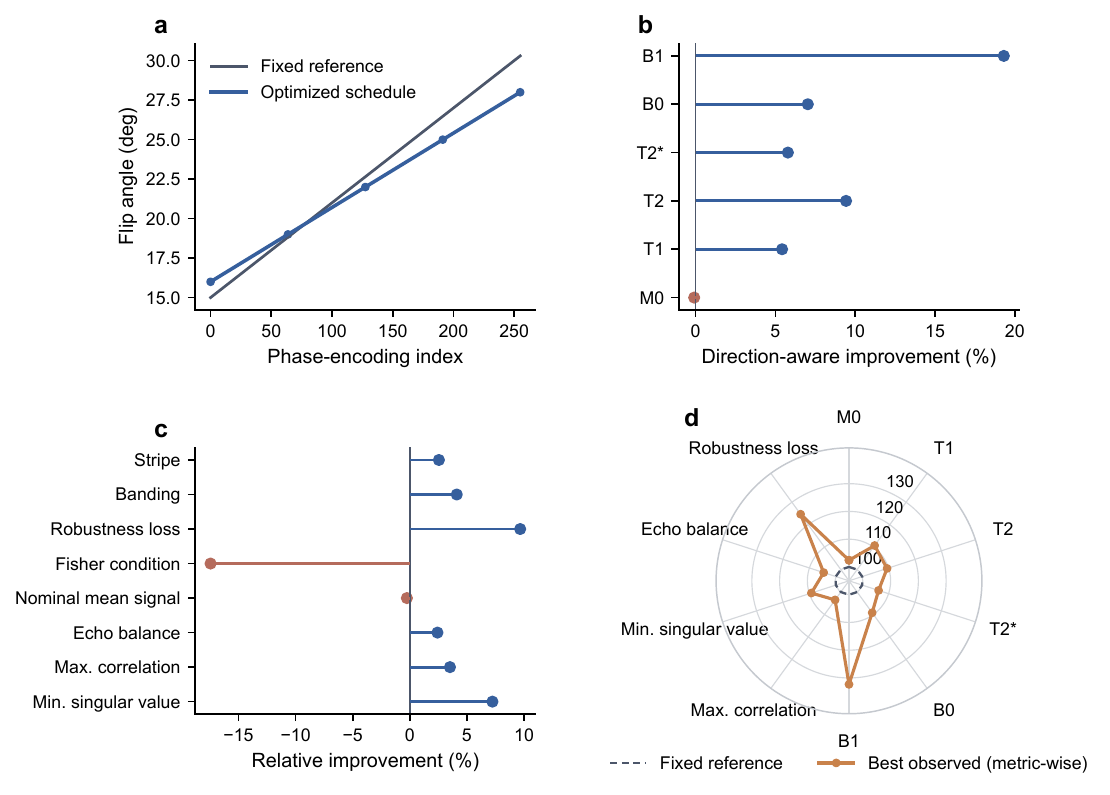}
\caption{Archive-selected candidate and multiobjective trade-offs in the original search. (a) Flip-angle trajectories for the fixed reference and candidate \#353; (b) direction-aligned changes in six CRLB-related standard-deviation proxies, with a slight degradation in $M_0$; (c) direction-aligned changes in other physical and diagnostic metrics, where negative values indicate adverse changes; and (d) the independently best level of each metric in the eligible archive, normalized to $100$ for the reference. Different axes may reflect different candidates; panel (d) is not the complete metric vector of candidate \#353.}
\label{fig:archive-metrics}
\end{figure}

\begin{table}[htbp]\centering\footnotesize
\caption{Multiobjective metrics for the fixed reference and archive-reselected candidate in the original search.}\label{tab:archived-metrics}
\begin{tabular}{@{}lrrr@{}}\toprule
Metric & Fixed reference & Selected candidate & Relative change (\%)\\\midrule
$r_{M_0}$ & 0.002310443 & 0.002312405 & +0.0849\\
$r_{T_1}$ & 0.006564837 & 0.006208974 & -5.4207\\
$r_{T_2}$ & 0.003369154 & 0.003051537 & -9.4272\\
$r_{T_2^*}$ & 0.005667921 & 0.005340107 & -5.7837\\
$r_{B_0}$ & 0.0001120633 & 0.0001041838 & -7.0313\\
$r_{B_1}$ & 0.001094844 & 0.0008834532 & -19.3078\\
Echo balance & 0.7411548 & 0.7590986 & +2.4211\\
Nominal full-image signal & 0.02143057 & 0.02137544 & -0.2572\\
Robustness loss & 0.03912206 & 0.03534090 & -9.6650\\
Jacobian condition ratio & 7253.052 & 8518.723 & +17.4502\\
Minimum singular value & 116.1722 & 124.5783 & +7.2358\\
Maximum correlation & 0.7918823 & 0.7640412 & -3.5158\\
Band energy & 0.6579541 & 0.6309003 & -4.1118\\
Stripe energy & 0.08560738 & 0.08342651 & -2.5475\\
Aggregate score & 91.593889 & 92.035636 & ---\\\bottomrule
\end{tabular}
\end{table}
The selected trajectory has five control points $[16,19,22,25,28]^\circ$, with search identifier \#353, and originated from an LLM proposal. Equally spaced control-point values produce a linear trajectory after PCHIP interpolation. The candidate with the highest raw aggregate score, $96.077768$, failed safeguards for signal, echo balance, and robustness and was therefore ineligible. The selected eligible candidate scored $92.035636$, reflecting the separation of score comparison and scientific eligibility. Figure~\ref{fig:archive-metrics} presents the selected trajectory and multiobjective results. Panels (b) and (c) show direction-aligned metric changes for this single selected candidate relative to the fixed reference. Panel (d) shows the best value of each metric \emph{achieved separately by different candidates} in the full eligible archive, not the simultaneous performance of any one sequence.

\subsection{Independent Computational Replay and Sensitivity to Research Preferences}
The archive-reselected candidate was independently recomputed for all $17$ scenarios using the same numerical backend. Scientific eligibility was unchanged, and all $13$ mandatory comparison metrics satisfied relative tolerance $10^{-3}$ or absolute tolerance $10^{-9}$. The largest recorded relative metric difference was $0.0499029\%$ for the condition ratio. The original selection score was $92.035636$, compared with $92.034327$ on replay.

An offline one-factor-at-a-time analysis of the same $191$ complete candidates examined selection sensitivity. Reducing the allowable condition-ratio degradation from $20\%$ to $10\%$ left $39$ eligible candidates and made the selected candidate ineligible. Increasing it to $25\%$ left $58$ eligible candidates and preserved the selection. Prohibiting any $M_0$ degradation left $15$ eligible candidates and changed the selection. Reducing the maximum-correlation weight from $0.15$ to $0.10$ preserved the eligible set but changed the top-ranked candidate. These outcomes characterize sensitivity of archive-based selection to research preference settings.

\FloatBarrier
\section{Extension-Task Validation for GRE and SE}
GRE and SE each used three random seeds (11, 23, 37) and two predefined configurations: numerical-only search (N) and agent-assisted without retrieval augmentation (A). GRE varied flip angle ($5$--$60^\circ$), TR (10--60 ms), and TE (2--12 ms); SE varied TR (300--1500 ms) and TE (20--120 ms), with excitation and refocusing flip angles fixed at $90^\circ/180^\circ$. Both task families evaluated worst-case two-tissue region-of-interest (ROI) contrast together with total acquisition time including preparation, using family-specific timing and scientific checks. Each task used staged evaluations of 16 candidates at $16^2$, 4 at $32^2$, and 1 at $64^2$ resolution (with five scenarios in each of the latter two stages), followed by a five-scenario replay of the selected candidate.
\begin{table}[tp]\centering\footnotesize
\caption{Per-configuration results for GRE extension tasks, assessing the contrast--time trade-off.}\label{tab:gre}
\begin{tabular}{@{}lrrrccl@{}}\toprule
Configuration/seed & FA ($^\circ$) & TR (ms) & TE (ms) & Contrast & Duration (s) & Outcome\\\midrule
N/11 & 20.000 & 30.000 & 5.000 & 0.02314551 & 7.920 & Eligible/replay passed\\
A/11 & 30.000 & 15.000 & 4.500 & 0.01387110 & 6.960 & Eligible/replay passed\\
N/23 & --- & --- & --- & --- & --- & Timing rejected\\
A/23 & 27.368 & 16.812 & 7.441 & 0.01480514 & 7.081 & Eligible/replay passed\\
N/37 & 44.058 & 58.026 & 6.566 & 0.03297491 & 9.747 & Eligible/replay passed\\
A/37 & 40.000 & 55.000 & 9.500 & 0.02924808 & 9.570 & Eligible/replay passed\\\bottomrule
\end{tabular}
\end{table}
\begin{table}[tp]\centering\footnotesize
\caption{Per-configuration SE extension results. The N/A configurations selected the same parameters within each seed.}\label{tab:se}
\begin{tabular}{@{}lrrccl@{}}\toprule
Configuration/seed & TR (ms) & TE (ms) & Contrast & Duration (s) & Outcome\\\midrule
N/11 & 550.153 & 68.168 & 0.02616991 & 41.262 & Eligible/replay passed\\
A/11 & 550.153 & 68.168 & 0.02616988 & 41.262 & Eligible/replay passed\\
N/23 & 669.522 & 71.939 & 0.03671176 & 48.875 & Eligible/replay passed\\
A/23 & 669.522 & 71.939 & 0.03671179 & 48.875 & Eligible/replay passed\\
N/37 & 931.679 & 82.381 & 0.06287580 & 66.149 & Eligible/replay passed\\
A/37 & 931.679 & 82.381 & 0.06287582 & 66.149 & Eligible/replay passed\\\bottomrule
\end{tabular}
\end{table}
GRE configuration N/23 terminated at high-resolution instantiation because its timing constraints were not met and did not yield an eligible final design. Replay passed for all other $11$ tasks, with a maximum recorded relative metric difference of $0.000234544\%$. These results illustrate the use of a common research workflow to coordinate candidate construction, physical evaluation, and replay under different parameterizations and constraints.

\section{Additional Experiments for FSE and Quantitative Optimization}
\subsection{FSE Design Identities and Fixed Cross-Phantom Verification}
A1--A3 in the main paper represent an actual agent feedback-and-revision chain. Fixed transfer and independent replay on three frozen phantoms used the original A3. The subsequent banded-order configurations centered at shots 0 and 16 further compared acquisition layouts with A3's timing parameters; their reconstructions are used in the main paper to show the effect of encoding order on spatial image characteristics.
\begin{table}[htbp]\centering\footnotesize
\caption{Fixed verification of original A3 on three frozen digital brain phantoms.}\label{tab:fse-three}
\begin{tabular}{@{}lrrrr@{}}\toprule Phantom & CSF/GM & GM/WM & Magnitude $L_2$ & Complex replay relative $L_2$\\\midrule
Brain01 & 0.44858 & 0.11669 & 0.48981 & $3.31\times10^{-6}$\\
Brain02 & 0.43341 & 0.11462 & 0.47585 & $2.37\times10^{-6}$\\
Brain03 & 0.42887 & 0.16109 & 0.49270 & $1.90\times10^{-6}$\\\bottomrule\end{tabular}
\end{table}
All three cases used ETL=8, ESP=25 ms, TR=3000 ms, $TE_{\rm eff}=125.004$ ms, and native $256\times256$ acquisition. Magnitude $L_2$ was calculated against the reference results for each phantom, whereas complex replay relative $L_2$ is reported separately as a repeatability metric. Outside-support energy for the banded layouts was $1.40180\%$ for original A3, $0.24577\%$ for shot 0, and $0.04083\%$ for shot 16. The corresponding CSF/GM values were $0.44858$, $0.45128$, and $0.50198$, and GM/WM values were $0.11669$, $0.07250$, and $0.06021$.

\subsection{Repeated Design Experiments for Four FSE Objectives}
The additional automated FSE design tests covered four predefined objectives: tissue contrast (FSE-A), scan time (FSE-B), RF resources (FSE-C), and spatial leakage (FSE-D). Three independent tasks were scheduled for each. All $12$ planned tasks produced scientifically eligible designs that passed independent replay. Including three original interruptions due to GPU occupancy, there were $15$ launch attempts and $12$ completions. Environment recovery followed a prespecified supplementary protocol that retained the original scientific thresholds and created new state records. First-candidate compilation succeeded in $11/12$ tasks; $12$ of $15$ search candidates completing native evaluation were scientifically eligible. All three tasks requiring scientific-feedback revisions produced eligible designs. Exactly $7/15$ launch attempts completed without initial environment repair or resource recovery. End-to-end times combine task records from two GPU platforms.

\begin{table}[htbp]\centering\footnotesize
\caption{Repeated design results and execution costs for four FSE research objectives.}\label{tab:fse-new}
\begin{tabular}{@{}lcccc@{}}\toprule
Objective & Eligible and replay passed & Mean candidates & Mean native evaluations & Mean end-to-end time (min)\\\midrule
Tissue contrast & 3/3 & 1.00 & 2.00 & 25.93\\
Scan time & 3/3 & 2.00 & 3.00 & 34.24\\
RF resources & 3/3 & 1.00 & 2.00 & 46.83\\
Spatial leakage & 3/3 & 1.00 & 2.00 & 70.87\\\bottomrule
\end{tabular}\end{table}

In one actual feedback chain for a scan-time task, the initial candidate failed retention and image-error criteria. A subsequent revision reduced relative magnitude $L_2$ from $0.18130$ to $0.08068$ and increased minimum tissue signal retention from $0.84351$ to $0.94530$, while scan time changed from $19.2$\,s to $38.4$\,s. A separate prespecified representative T1 design also passed scientific checks and replay on two independent numerical phantom configurations, Brain02 and Brain03. All tests were automated research tasks in synthetic phantom environments.

\subsection{$T_2$-Specific Study of Candidate \#186 and Paired Statistics}
Search identifier \#186 corresponds to Optuna NSGA-II trial 244 from the hybrid multiobjective search described in S3. Its five PCHIP control points were $[34.0141,42.1424,52.3014,49.9391,48.3275]^\circ$, forming a nonlinear flip-angle trajectory over $256$ phase-encoding lines. Under the original scientific safeguards, the candidate passed absolute execution checks but failed some reference-relative safeguards for $B_0$ sensitivity, echo balance, nominal signal, and worst-scenario robustness. After the original search, this candidate was selected for a subsequent $T_2$-focused imaging and reconstruction study.

Recomputation from stored features gave the normalized $T_2$ standard-deviation-bound proxy $r_{T_2}=0.00184195746$, compared with $0.00336915372$ for the fixed reference, a $45.3288\%$ decrease. Its observations, finite differences, magnitude normalization, and singular value decomposition follow the common archived evaluator in S2. The $T_2$-specific ranking considered candidates that passed absolute checks and had complete $T_2$ comparison data, whereas multiobjective archive selection required all scientific eligibility criteria. Figures S1 and S2 depict different evaluation aspects of the original multiobjective candidates.

The reference and $T_2$-specific candidate each generated $512$ paired synthetic Full-Bloch samples with identical $384/64/64$ training, validation, and test partitions. Complex five-echo observations were split into ten real and imaginary input channels, and separate R2AttUNet reconstructors were trained for each sequence. For the fixed $64$ synthetic test samples, MAE decreased from $5.690097$\,ms to $5.195594$\,ms. Mean paired improvement was $0.494503$\,ms, with median improvement $0.454307$\,ms. MAE improved in $60/64$ cases and masked SSIM in $64/64$; MAE increased in the other four cases. Leave-one-sample-out mean improvements ranged from $0.437312$ to $0.515700$\,ms. A one-to-one mapping between data-generation sources and independent subjects has not been established, so these statistics describe fixed paired synthetic samples. The ratio of sums of squared flip angles between the optimized and reference sequences was $4.124192$.

\section{Additional Numerical Validation of the SmartBloch Simulation Engine}
Single-voxel accuracy testing of SmartBloch CUDA covered 12 condition classes each for WM, GM, and CSF, totaling 36 classes and 288 complex echo results. An independent double-precision affine Bloch matrix-exponential calculation in NumPy/SciPy was used as the reference. Maximum complex absolute error for 27 instantaneous-RF classes (216 echoes) was $1.6924\times10^{-6}$ (threshold $2\times10^{-5}$); for 9 finite-duration-RF classes (72 echoes), it was $5.1256\times10^{-5}$ (threshold $2\times10^{-4}$).

Additional numerical tests of SmartBloch included 11 slice-selection conditions with gradients and the actual import pathway (352 complex ADC samples), yielding maximum complex error $4.91491\times10^{-5}$; one out-of-slice signal ratio did not satisfy its scientific eligibility criterion. Subsequent combined crusher and spatial tests covered 18 conditions and 576 ADC samples, with maximum complex error $2.62574\times10^{-5}$. A comparison of 65 versus 129 spatial nodes yielded complex relative $L_2$ of $3.69963\times10^{-5}$.

Main Section 4.3 reports runtime measurements for the current SmartBloch CUDA engine on GRE and GRE-EPI sequences, alongside comparative timings for MRiLab and SPROM.

\section{Agent Implementation and Execution Configuration}
\subsection{Model Calls and Records}
FSE design and SSFP-MOLED optimization both used DeepSeek V4 Flash, identified as \texttt{deepseek-v4-flash}, through DeepSeek's OpenAI-compatible function-calling API. Model calls used default API parameters without explicitly overriding temperature, top-$p$, maximum output length, stop tokens, or model seed. Fixing a task seed therefore does not make LLM candidate generation fully deterministic.

\begin{table}[htbp]\centering\footnotesize
\caption{Agent model-call configuration.}\label{tab:agent-call-config}
\setlength{\tabcolsep}{5pt}\renewcommand{\arraystretch}{1.12}
\begin{tabular}{@{}>{\raggedright\arraybackslash}p{4.0cm}>{\raggedright\arraybackslash}p{4.4cm}>{\raggedright\arraybackslash}p{4.4cm}@{}}\toprule
Setting & FSE design & SSFP-MOLED optimization\\\midrule
Model and interface & \texttt{deepseek-v4-flash}; OpenAI-compatible function calling & \texttt{deepseek-v4-flash}; OpenAI-compatible function calling\\
Model parameters & API defaults, no additional overrides & API defaults, no additional overrides\\
Model retries / tool rounds & 1 retry / 1 round & 2 retries / 2 rounds\\
Prompt and response records & Full prompts and raw responses & Prompt hashes and raw responses\\
Numerical random seeds & Fixed task seeds for Python, NumPy, and CUDA (2026101901--2026101903) & Fixed Optuna sampler seed 6303 for numerical search\\
\bottomrule
\end{tabular}
\end{table}

The shared agent system prompt restricts generative decisions to registered tools and the frozen scientific protocol:

\begin{verbatim}
You are the SmartBloch MRI sequence optimization controller.
Propose bounded candidates and use registered tools for experiment actions.
Never invent metrics, alter completed evidence, or exceed absolute bounds.
Selection never implies completion; deterministic scientific gates and replay
validation determine whether an experiment is complete.
\end{verbatim}

Candidate parameters must originate from the registered schema and remain within frozen bounds. Feedback-driven revisions must reference actual parent candidates and feedback events. Final selection must draw from the permitted candidate set. FSE task prompts additionally fix native $256\times256$ physical models and objective-specific design bounds and disallow numerical-search fallback. SSFP-MOLED prompts fix five-echo timing, tissue models, and flip-angle parameterization, with the agent coordinating candidate exploration from stage feedback.

\subsection{Task Inputs, Tools, and Budgets}
Model inputs include task objectives, current state, parameter schemas and bounds, metric directions and eligibility rules, evaluated candidates, feedback events, and candidate lineage. Outputs are structured JSON decisions containing action types, supporting evidence identifiers, candidate parameters, anticipated effects, and diagnostic information. Repeated FSE design tasks expose only \texttt{enqueue\_candidates}. Formal SSFP-MOLED tasks expose \texttt{continue\_optimization}, \texttt{enqueue\_candidates}, and \texttt{select\_final\_candidate}; Optuna/NSGA-II performs continuous numerical search.

\begin{table}[htbp]\centering\footnotesize
\caption{Execution budgets for the two principal experiments.}\label{tab:agent-budgets}
\setlength{\tabcolsep}{4pt}\renewcommand{\arraystretch}{1.12}
\begin{tabular}{@{}>{\raggedright\arraybackslash}p{3.5cm}>{\raggedright\arraybackslash}p{4.8cm}>{\raggedright\arraybackslash}p{4.8cm}@{}}\toprule
Budget item & Repeated FSE design & SSFP-MOLED\\\midrule
Native search candidate budget & At most 4 initial candidates per task & 200 candidates\\
Compile-only rejection budget & At most 8 & Counted toward total candidate budget\\
Model decision budget & 16 & 200\\
Model retries / tool rounds & 1 retry / 1 round & 2 retries / 2 rounds\\
Candidate retries & 0 & 1\\
Independent final replay & 1 & 1\\
Native execution timeout & 7200\,s & 43200\,s\\
\bottomrule
\end{tabular}
\end{table}

After parameter validation, Pulseq compilation, native Full-Bloch computation, acquisition reconstruction, and fixed scientific eligibility checks, FSE candidates advance to the next iteration based on actual feedback; numerical-search fallback is not used. For SSFP-MOLED, the agent interprets stage results and changes exploration direction, while numerical optimizers generate candidates. Only candidates passing execution checks and scientific eligibility enter Pareto comparison. Both task types stop when budgets are exhausted, no eligible candidate is found, or resource-management conditions terminate execution; final candidates must pass independent replay. Malformed or empty model responses are retried within the model budget. Unknown tools, invalid parameters, and out-of-bounds proposals are rejected, with compilation failures and scientific ineligibility recorded separately.

\section{Quantitative Workflow Comparison and Timing Definitions}
\subsection{Per-Objective Timing for FSE Tasks}
Four FSE design objectives were considered: tissue contrast, scan time, RF resources, and spatial leakage. Table~\ref{tab:fse-workflow-time-detail} summarizes execution time across the corresponding workflows. The time accounting incorporates objective-specific parameter adjustment, scientific-feedback interpretation, design revision, and workflow coordination.

\begin{table}[htbp]\centering\footnotesize
\caption{Execution time across four FSE design objectives.}\label{tab:fse-workflow-time-detail}
\begin{tabular}{@{}lrrr@{}}\toprule
Research objective & SmartBloch-Agent (min) & Fixed workflow (min) & Manual coordination (min)\\\midrule
Tissue contrast & 25.93 & $\approx43$ & $\approx82$\\
Scan time & 34.24 & $\approx59$ & $\approx126$\\
RF resources & 46.83 & $\approx87$ & $\approx179$\\
Spatial leakage & 70.87 & $\approx129$ & $\approx274$\\
\textbf{Mean} & \textbf{44.47} & \textbf{$\approx80$} & \textbf{$\approx165$}\\\bottomrule
\end{tabular}
\par\smallskip
\begin{minipage}{0.98\linewidth}\raggedright\footnotesize
Execution time covers sequence construction, physical evaluation, feedback-driven refinement, and workflow coordination. Fixed and manually coordinated workflows use task-specific time budgets reflecting differences in scientific objectives, parameter adjustment, and sequence-design complexity. Values are summarized per research task.
\end{minipage}
\end{table}
\FloatBarrier

The mean end-to-end times were $44.47$\,min for SmartBloch-Agent, approximately $80$\,min for the fixed workflow, and approximately $165$\,min for manual coordination. The corresponding human-effort budgets were approximately $25$ and $100$\,min per task for the fixed and manually coordinated workflows, respectively. Coordination requirements varied with the scientific objectives, with spatial-leakage refinement involving additional acquisition-order analysis and reconstruction-based assessment.

For tissue contrast, scan time, RF resources, and spatial leakage, fixed-workflow human-effort budgets were $12$, $18$, $27$, and $40$\,min, respectively; the corresponding manual-coordination budgets were $48$, $75$, $105$, and $160$\,min. Additional review allowances were $5$, $7$, $13$, and $18$\,min for the fixed workflow and $8$, $17$, $27$, and $43$\,min for manual coordination. The alternative-workflow end-to-end budgets combine the corresponding SmartBloch-Agent execution time, human-effort budget, and review allowance, rounded to the nearest minute. These are workload budgets rather than separately recorded execution times.

\subsection{Simulation and Coordination Costs for SSFP-MOLED}
The full SSFP-MOLED search completed $1{,}135$ scenarios. Agent-recorded stage times are given in Table~\ref{tab:ssfp-stage-time}. F2 accounted for approximately $99.3\%$ of simulation and evaluation time, so end-to-end cost was dominated by multiecho physical simulation and high-fidelity evaluation.

\begin{table}[htbp]\centering\footnotesize
\caption{Stage-wise execution times for the SSFP-MOLED agent search.}\label{tab:ssfp-stage-time}
\begin{tabular}{@{}lrrr@{}}\toprule
Stage & Scenarios & Cumulative time (h) & Mean time per scenario (s)\\\midrule
F0 & 192 & 0.205 & 3.84\\
F1 & 736 & 1.431 & 7.00\\
F2 & 207 & 296.626 & 5158.71\\
Recorded total & 1135 & 298.707 & 946.03\\\bottomrule
\end{tabular}
\par\smallskip
\begin{minipage}{0.96\linewidth}
\raggedright\footnotesize Note: Cumulative F0, F1, and F2 time was $298.262$\,h. The recorded total of $298.707$\,h additionally includes $0.445$\,h for one run recovery. Mean time per scenario is calculated from cumulative time within each of the three stages and excludes recovery overhead.
\end{minipage}
\end{table}

\begin{table}[htbp]\centering\footnotesize
\caption{Human effort across SSFP-MOLED research activities.}\label{tab:ssfp-human-effort}
\begin{tabularx}{\linewidth}{@{}Xrr@{}}\toprule
Research activity & Fixed workflow (h) & Manual coordination (h)\\\midrule
Research planning & $\approx2$ & $\approx5$\\
Feedback analysis and strategy adjustment & $\approx6$ & $\approx25$\\
Final selection and verification & $\approx2$ & $\approx6$\\
\textbf{Total} & \textbf{$\approx10$} & \textbf{$\approx36$}\\\bottomrule
\end{tabularx}
\par\smallskip
\begin{minipage}{0.98\linewidth}\raggedright\footnotesize
Research planning covers objective specification and evaluation configuration. Feedback analysis includes candidate assessment, multiobjective interpretation, scientific-eligibility checks, and search-strategy adjustment. Final selection includes candidate ranking, evidence review, and verification. Human effort is accumulated over the complete optimization study.
\end{minipage}
\end{table}

The SSFP-MOLED optimization comprised $1{,}135$ evaluation scenarios across three fidelity stages. As summarized in Table~\ref{tab:ssfp-stage-time}, the F0, F1, and F2 stages required $0.205$, $1.431$, and $296.626$\,h, respectively. High-fidelity physical evaluation accounted for approximately $99.3\%$ of the total recorded execution time.

Research coordination comprised objective specification, candidate assessment, scientific-feedback interpretation, strategy adjustment, and final selection. Table~\ref{tab:ssfp-human-effort} summarizes the associated human-effort budgets. Fixed workflow coordination required approximately $10$\,h, while manual coordination required approximately $36$\,h. Feedback analysis and strategy adjustment constituted the largest share of human involvement. SmartBloch-Agent integrated these research operations into an autonomous workflow, coordinating physical evaluation and scientific decision-making without manual scientific-parameter editing. The fixed and manual figures are coordination-time budgets for the same evaluation workload; physical-evaluation time is reported separately in Table~\ref{tab:ssfp-stage-time}.

\section{Data and Training Configuration for \texorpdfstring{$T_2$}{T2} Quantitative Reconstruction}
The reference and $T_2$-specific sequences each generated $512$ synthetic Full-Bloch samples, matched for sample identity, tissue parameter ground truth, valid masks, and the $384/64/64$ training, validation, and test partitions. Samples were selected from $6480$ source \texttt{VObj} files. Source data were $512\times512$ and resampled to $256\times256$. Parameter ranges in the source pool were $M_0=0.05$--$1.0668$, $T_1=0.0954$--$5.7440$\,s, $T_2=4.25$--$406.93$\,ms, $T_2^*=2.55$--$390.18$\,ms, $B_0=-200$--$200$\,Hz, and $B_1=0.7$--$1.5$. Five-echo complex data were separated into ten real and imaginary input channels, with a $T_2$ map as output. Valid masks required positive $M_0$, $T_1$, $T_2$, $T_2^*$, and $B_1$, together with $T_2^*\leq T_2$. No additional measurement noise was introduced in formal training or testing, and spatial $B_0/B_1$ variation in the source data was retained. The $\rho$-morphology descriptor identified $117$ morphology groups for partitioning; detected morphology matches did not cross partitions. The $384$ training samples came from one morphology group, whereas validation and test sets covered $52$ and $53$ groups, respectively.

\begin{table}[htbp]\centering\footnotesize
\caption{$T_2$ reconstruction network and training settings.}\label{tab:t2-training-config}
\setlength{\tabcolsep}{5pt}\renewcommand{\arraystretch}{1.12}
\begin{tabular}{@{}p{4.0cm}p{9.4cm}@{}}\toprule
Item & Setting\\\midrule
Network & R2AttUNet; 10 input channels, 1 output channel; 16 base filters; 2 recurrences; 1,485,709 parameters\\
Output and loss & Softplus output; foreground masked L1 $+\,0.05$ times background masked L1\\
Optimizer & AdamW; learning rate $10^{-3}$; weight decay $10^{-5}$; batch size 4\\
Training procedure & Up to 150 epochs; checkpoint selected by validation MAE; early stopping after 25 epochs without improvement\\
Learning-rate schedule and gradients & ReduceLROnPlateau, factor 0.5, patience 8; gradient clipping 5\\
Reproducibility setting & Seed 20261014; FP32; RTX 2080 Ti; PyTorch 2.1.2+cu118\\
Normalization & Separate complex RMS normalization on the training sets for reference and $T_2$-specific sequences, with scales 0.03283059 and 0.04478420\\
\bottomrule
\end{tabular}
\end{table}

Each network was trained only on complex observations generated by its corresponding sequence, and checkpoints were selected only by validation MAE. The test set was reserved for final evaluation. Results and paired distributions are reported in the main paper's paired reconstruction table and Supplementary Section S5.

\section{Sequence Output Packages and Research Evidence}
The final FSE design output package contains the following files:
\begin{itemize}[leftmargin=2em,itemsep=1pt,topsep=3pt]
\item \texttt{final\_optimized\_sequence.seq}: executable sequence file;
\item \texttt{final\_sequence\_spec.json}: complete event, timing, and key design parameters;
\item \texttt{final\_optimized\_config.json}: final runtime configuration;
\item \texttt{evidence\_bundle.json}: final candidate, independent replay results, and file SHA-256 hashes.
\end{itemize}
The task directory also stores task and scientific contracts, candidate lineage, LLM decision records, timing and event lists, complex samples, reconstructed images, scientific metrics, execution logs, and hardware execution information.

Native SSFP-MOLED outputs include complex $k$-space, complex reconstructions, magnitude and phase images, and readout trajectories.

The metric file is \texttt{native\_metrics.json}; the file manifest records data shapes, types, axis definitions, units, and hashes.

Pulseq sequence files can be included as exchange files in the output package. Actual scanner acquisition requires a corresponding Pulseq interpreter, vendor-specific sequence environment, and on-site hardware safety checks.

\section{Reproduction Scope and Archive Information}
Candidate counts and screening outcomes of the original hybrid search are documented in S3. Optuna candidate \#186 and LLM candidate \#353 originated from the same search archive; the subsequent $T_2$ quantitative reconstruction used separately generated paired data and training procedures.

\nocite{yang2026ssfpmoled,layton2017pulseq,akiba2019optuna,deb2002nsgaii}\bibliographystyle{unsrt}\bibliography{references}